%% file: main.tex
\documentclass[twoside,twocolumn,9pt]{article}
\usepackage{extsizes}
\usepackage[super,sort&compress,comma]{natbib} 
\usepackage[version=4]{mhchem}
\usepackage[left=1.5cm, right=1.5cm, top=1.785cm, bottom=2.0cm]{geometry}

\usepackage{balance}
\usepackage{widetext}
\usepackage{times,mathptmx}
\usepackage{sectsty}
\usepackage{graphicx} 
\usepackage{lastpage}
\usepackage[format=plain,justification=raggedright,singlelinecheck=false,font={stretch=1.125,small,sf},labelfont=bf,labelsep=space]{caption}
\usepackage{float}
\usepackage{afterpage}
\usepackage{fancyhdr}
\usepackage{fnpos}
\usepackage[english]{babel}
\usepackage{array}
\usepackage{droidsans}
\usepackage{charter}
\usepackage[T1]{fontenc}
\usepackage[usenames,dvipsnames]{xcolor}
\usepackage{setspace}
\usepackage[compact]{titlesec}
\usepackage{color}
\usepackage{xspace}
\usepackage{comment}
\usepackage{siunitx}
\usepackage{soul}
\usepackage{chemformula} 

\usepackage{multirow}
\usepackage{subfig}

\definecolor{cream}{RGB}{222,217,201}

\newcommand{\beginsupplement}{%
        \setcounter{table}{0}
        \renewcommand{\thetable}{S\arabic{table}}%
        \setcounter{figure}{0}
        \renewcommand{\thefigure}{S\arabic{figure}}%
        \setcounter{figure}{0}
        \renewcommand{\thesubsection}{S\arabic{subsection}}%
     }

\usepackage{booktabs}
\usepackage{multirow}
\usepackage{stfloats}

\DeclareSIUnit{\rpm}{rpm}
\DeclareSIUnit{\dB}{dB}
\DeclareSIUnit{\ppm}{ppm}
\DeclareSIUnit{\ppb}{ppb}
\DeclareSIUnit{\Molar}{M}

\begin{document}

\makeatletter 
\newlength{\figrulesep} 
\setlength{\figrulesep}{0.5\textfloatsep} 

\newcommand{\topfigrule}{\vspace*{-1pt}%
\noindent{\color{cream}\rule[-\figrulesep]{\columnwidth}{1.5pt}} }

\newcommand{\botfigrule}{\vspace*{-2pt}%
\noindent{\color{cream}\rule[\figrulesep]{\columnwidth}{1.5pt}} }

\newcommand{\dblfigrule}{\vspace*{-1pt}%
\noindent{\color{cream}\rule[-\figrulesep]{\textwidth}{1.5pt}} }

\makeatother

\twocolumn[
  \begin{@twocolumnfalse}
 
\vspace{1cm}  
\sffamily
\begin{tabular}{p{18cm} }

\noindent\LARGE{\textbf{An NMR-compatible microfluidic platform enabling \textit{in~situ} electrochemistry$^\dag$} }\\

\vspace{0.3cm} \\
\noindent\large{Hossein Davoodi{\textit{$^{a}$}}, Nurdiana Nordin{\textit{$^{a,b}$}}, Lorenzo Bordonali{\textit{$^{a}$}}, Jan G. Korvink{\textit{$^{a}$}}, Neil MacKinnon{\textit{$^{a,*}$}}, and Vlad Badilita{\textit{$^{a,*}$}}} \\

\vspace{0.5cm} \\

\normalsize{\noindent
Combining microfluidic devices with nuclear magnetic resonance (NMR) has the potential of unlocking their vast sample handling and processing operation space for use with the powerful analytics provided by NMR.  One particularly challenging class of integrated functional elements from the perspective of NMR are conductive structures. Metallic electrodes could be used for electrochemical sample interaction for example, yet they can cause severe NMR spectral degradation. In this study, a combination of simulation and experimental validation was used to identify an electrode geometry that, in terms of NMR spectral parameters, performs as well as for the case when no electrodes are present. By placing the metal tracks in the side-walls of a microfluidic channel, we found that NMR RF excitation performance was actually \textit{enhanced}, without compromising \ce{B0} homogeneity.  Monitoring \textit{in~situ} deposition of chitosan in the microfluidic platform is presented as a proof-of-concept demonstration of NMR characterisation of an electrochemical process. 
}
\end{tabular}

 \end{@twocolumnfalse} \vspace{0.6cm}

 ]

\renewcommand*\rmdefault{bch}\normalfont\upshape
\rmfamily
\section*{}
\vspace{-1cm}

\footnotetext{\textit{$^{a}$Institute of Microstructure Technology (IMT), Karlsruhe Institute of Technology (KIT), Hermann-von-Helmholtz Platz 1, 76344 Eggenstein-Leopoldshafen, Germany}}
\footnotetext{\textit{$^{b}$Department of Chemistry, Faculty of Science, University of Malaya, 50603 Kuala Lumpur, Malaysia}}
\footnotetext{\textit{$^{*}$Corresponding authors: NM (neil.mackinnon@kit.edu), VB (vlad.badilita@kit.edu)}}
\footnotetext{\dag~Electronic Supplementary Information (ESI) available. See DOI: 00.0000/00000000.}

\input{01_Introduction.tex}

\input{02_Methods.tex}

\input{03_Results.tex}

\input{04_Conclusions.tex}

\input{05_Acknowledgements.tex}

\bibliographystyle{rsc} 
\bibliography{main}
\newpage
\onecolumn

\section*{\centering Electronic Supplementary Information}
\input{06_ElectronicSupplementaryInfo.tex}

\end{document}

%% file: 01_Introduction.tex
\section{Introduction}

The advances presented in this paper deal with the challenges arising at the intersection of three general fields of research: nuclear magnetic resonance (NMR), electrochemistry (EC), and lab-on-a-chip (LOC) techniques and devices. NMR offers unparalleled chemical specificity, being able to simultaneously identify and quantify hundreds of different compounds in complex mixtures. Moreover, NMR can access an extraordinary range of characteristic times for the processes under study, spanning more than 10 orders of magnitude, from  \SI{1e-9}{\second} to \SI{1e2}{\second}. NMR provides information from all states of matter and in a wide range of temperatures, and is able to \textit{see} inside the external boundaries of a sample in a non-destructive manner, being therefore the method of choice in situations where other spectroscopic methods fail.

All these beautiful characteristics make NMR an extremely interesting method with which to investigate the bio-chemical processes and phenomena usually studied within LOC devices. However, scientists relying on NMR to characterise processes in a LOC device must be aware of the sword of Damocles potentially endangering their results: NMR has a notoriously low intrinsic mass sensitivity. As an order of magnitude, consider protons in a \SI{9.4}{\tesla} magnetic field at room temperature, where only 10 out of one million spins contribute to the NMR signal, so that concentrations below \SI{1}{\milli\Molar} are hard to observe.

One proven way to address the issue of poor mass sensitivity in small samples is to resort to miniaturised NMR detectors.\cite{Badilita2012} An additional advantage is that many micro-MR topologies can be readily integrated with microfluidics. The straightforward geometry and fabrication process of planar coils have been employed to demonstrate multinuclear single and double resonance NMR in a microfluidic detection volume of \SI{25}{\nano \litre}.\cite{Fratila2014} Boero \textit{et al.} \cite{Grisi2018} successfully demonstrated NMR experiments on sub-nL samples, i.e. ova of microorganisms, by combining highly sensitive single-chip CMOS detectors with a high spatial resolution 3D printed microfluidic channel. Microcoil phased arrays combine the advantage of increased signal-to-noise ratio with an enlarged field of view,\cite{Gruschke2012} offering a \SI{2.5}{D} sensitive region that may include microfludic networks or sample chambers. The stripline detector topologies employed by Kentgens \textit{et al.}\cite{VanMeerten2018, Oosthoek-deVries2017} and Utz \textit{et al.}\cite{Hale2018, Sharma2019} have proven to be easily amenable to microfluidic integration, at the same time providing excellent \ce{B1} and \ce{B0} homogeneity. Solenoidal microcoils \cite{Badilita2010} must be integrated with three-dimensional microfluidic networks \cite{Meier2014} for sample delivery, which makes the fabrication process rather challenging. Another NMR detector which naturally lends itself to facile microfluidic integration is the Helmholtz coil geometry.\cite{Spengler2014, Spengler2016} Microfabrication technologies in general, and microfluidics in particular, offer several other valuable additions to NMR characterisation. Bordonali \textit{et al.} and Eills \textit{et al.} have indpendently reported microfluidic solutions which provide bubble-free hydrogen gas contact with the liquid sample, thus enabling the use of signal enhancement by means of parahydrogen induced hyperpolarisation at the microscale.\cite{Bordonali2019,Eills_High_2019} Similarly, an earlier paper reports on a 3D-printed system to optimise the dissolution of hyperpolarised gaseous species for microscale NMR and the functionality is demonstrated using \textsuperscript{129}Xe.\cite{Causier2015} 

Planar microfabrication including thin layer deposition of metals, UV-photolithography and etching allow for facile integration of electrodes in the microfluidic system, thus enabling an entire different class of experiments to be coupled with NMR, within a microfluidic environment for small amounts of samples. In a recent paper, spatial and temporal control over multilayer assembly and composition of chitosan hydrogel, by means of electrodeposition in a microfluidic environment, was demonstrated.\cite{Nordin2019}

There are several application scenarios that would benefit from the introduction of metallic electrodes in, or in the vicinity, of the NMR detection region. These scenarios include NMR hyphenated with electrokinetic separation methods (e.g., capillary electrophoresis), dielectrophoresis, or a wide range of electrochemical experiments. The presence of a metallic electrode structure impacts the NMR measurements in two ways: 1) it affects the \ce{B0} homogeneity due to a magnetic susceptibility jump at the leading edges of the tracks, with direct implications on the spectral resolution, and 2) it affects the \ce{B1} strength and homogeneity due to radiofrequency shielding effects, with direct implications on the sensitivity and uniformity of the NMR measurement. Metal track susceptibility is a rather old issue which has been reported during the very early EC-NMR attempts.\cite{Richards1975} Potential solutions to the \ce{B0} homogeneity problem, which stem from the Gauss law, have been identified in the re-design of metal electrodes with a high degree of symmetry \cite{Mincey1990,Prenzler2000}, or rotation of the sample in a similar way as is done in solid-state NMR.\cite{Richards1975,Prenzler2000,Albert1987,Mairanovsky1983} The second challenge in many ways is an orthogonal problem, with a different physical origin, namely stemming from the Ampere's law, and therefore requiring a different strategy. However, the solution for both problems is, to some extent, geometrical, so that neither the \ce{B0} nor the \ce{B1} field should intercept the largest projected surface of the metal electrodes. 
When the \ce{B1} field lines penetrate through a metal layer, the amplitude of the RF pulse is attenuated due to the skin effect. The characteristic quantity governing this effect is the skin depth, i.e., the material thickness which attenuates the RF amplitude by 1/e. The skin depth varies as $1/\sqrt{\sigma \times \omega}$, depending on the conductivity $\sigma$ of the electrode material, as well as on the frequency $\omega$ of the electromagnetic field (assuming non-magnetic electrode materials). Therefore, higher \ce{B0} fields and thus higher Larmor frequencies admit thinner metals as EC electrodes, which is an additional limitation for EC applications requiring high applied currents. Addressing this issue, some authors \cite{Prenzler2000, Webster2004} have employed electrodes with nanometer thickness, i.e., less than $1\%$ of the skin depth. Chen \textit{et al.} provided an analytical analysis of the field penetration inside a metallic cylinder.\cite{Chen2017} According to their study, metallic layers thinner than \SI{0.1}{\%} of the skin depth are transparent to the external RF field. In microsystems such as lab-on-a-chip devices, \ce{B0} and \ce{B1} perturbations are further exacerbated by the inherently smaller available real-estate.\cite{Ryan2014}

In this paper, acknowledging the important potential of NMR to perform \textit{in~situ} characterisation and monitoring of electrochemical reactions, we address the issues that arise by introducing metallic electrodes in the proximity of the NMR detection region, in particular for the case when lab-on-a-chip devices, miniaturised NMR detectors, and small (sub-\si{\micro\liter}) sample quantities are involved. We conjecture that this work is relevant for the $\mu$-TAS community, because the theoretical considerations derived and experimental characterisations presented herewith are also applicable to other NMR-hyphenated scenarios where metallic structures must be similarly introduced, i.e., in electrophoresis and dielectrophoresis, and digital microfluidics based on electrowetting, in which individual droplets are manipulated on a surface of an array of electrodes by means of individually applied voltages.\cite{Swyer_2016, Swyer_2019, Wu_2019}

\begin{figure*}
    \centering
    \includegraphics[width=\linewidth]{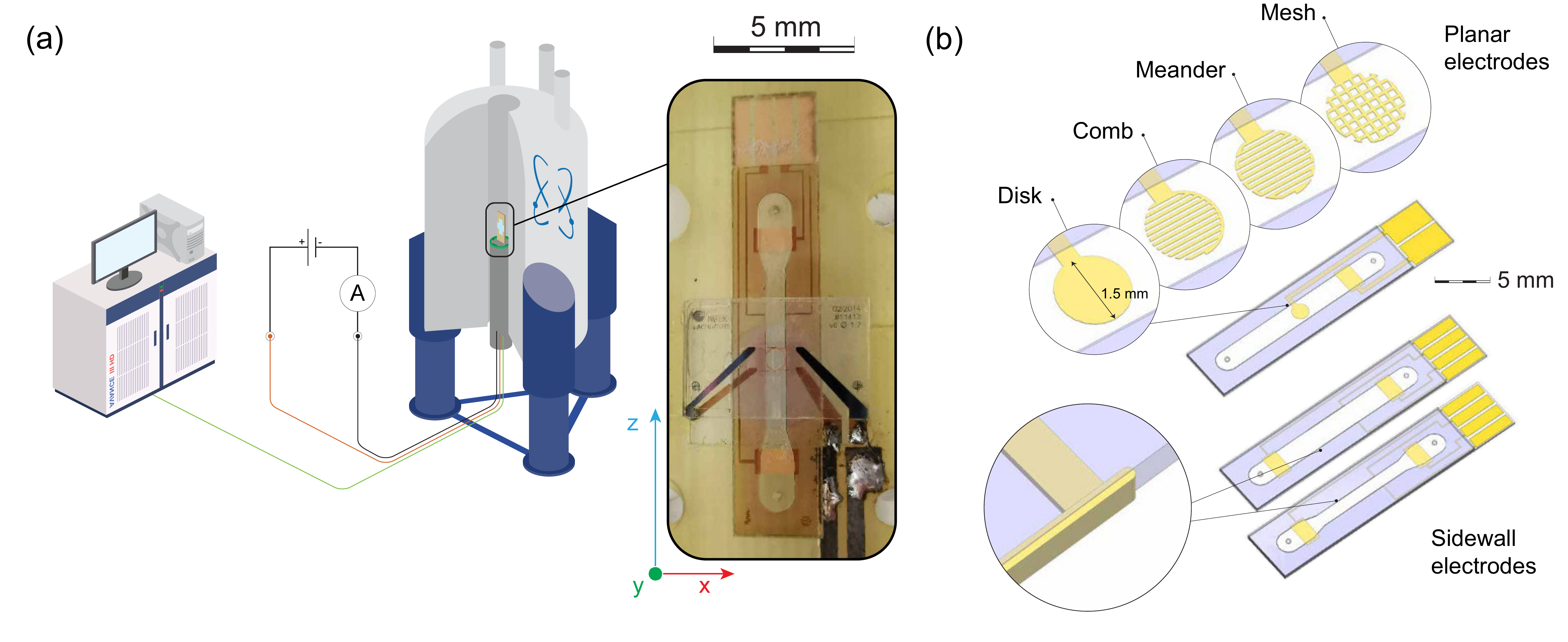}
    \caption{(a) Illustration of the setup for \textit{in-situ} electrochemistry using a commercial superconductive NMR/MRI system equipped with a micro Helmholtz coil, being tuned and matched at the \textsuperscript{1}H Larmor frequency, and a sample insert containing metallic electrodes. (b) Different sample inserts employing different electrode configurations, i.e, a planar active electrode (a disk, a comb, a meander, and a mesh geometry), narrow-channel sidewall electrodes, wide-channel sidewall electrodes.   
    }
    \label{fig:setup}
\end{figure*}

As a case study, we consider a microfluidic channel oriented along the static magnetic field \ce{B0} and various electrode topologies within this channel. The miniaturised NMR detector is a Helmholtz coil introduced previously by Spengler \textit{et al.}\cite{Spengler2016} In this way, we gain complete freedom to design a suitable microfluidic insert and the distribution of electrodes, the only geometrical precondition being that the insert fits inside the Helmholtz structure, as shown in Figure~\ref{fig:setup}a.

Various detector geometries were considered theoretically and a selection of these variants have been fabricated and tested. The study focuses on the effect of metal electrodes on the static field \ce{B0}, RF field \ce{B1}, and the additional noise captured in the detector. As a test platform to explore the practical effects of electrode integration with NMR-compatible microfluidics, we monitor the \textit{in~situ} electrodeposition of chitosan inside the microfludic channel. Chitosan (CS) is a biocompatible polyaminosaccaharide with the ability to undergo reversible gelation in response to a change in solution \si{pH}.\cite{cheng2010} Chitosan plays the role of forming an interface between the technical device (here, the NMR detector) and some biological sample/process.\cite{li2018, zhuang2018, chung2018} The electrodeposition process and the change in solution \si{pH} are monitored by NMR spectroscopy.

%% file: 02_Methods.tex
\section{Materials and methods}

\subsection{Chemicals and materials}

The following chemicals were purchased from Sigma Aldrich: sucrose, 3-(trimethylsilyl)- propionic-2,2,3,3-d4 acid sodium salt (TSP), sodium chloride (\ce{NaCl}), copper sulphate (\ce{CuSO4 $\cdot$ 5H2O}), chitosan from crab shells (\SI{85}{\%} deacetylation, 200kDa), poly(ethylene glycol) bis(carboxymethyl) ether (average M$_n$ 250, PEG), N-(3-(dimethylamino)propyl)-N-ethylcarbodiimide hydrochloride (EDC) \SI{99}{\%}, N-hydroxy-succinimide (\ce{NHS}) \SI{98}{\%}, deuterated chloride (\ce{DCl})(\SI{99}{\%} D), deuterated sodium hydroxide (\ce{NaOD}) (\SI{99.5}{\%} D) and deuterium oxide \ce{D2O} (\SI{99.9}{\%} D). 

\subsection{Simulation}
COMSOL Multiphysics\textregistered (COMSOL AB, Sweden) was used to perform FEM simulations for both RF (RF Module) and susceptibility (AC/DC module) studies. For this purpose, each insert geometry was simulated together with the Helmholtz pair. The relative tolerances of the simulations were set to \SI{1e-3}{} and \SI{1e-9}{} for RF and susceptibility simulations, respectively.
The simulation results were post-processed in MATLAB\textregistered (MathWorks Inc., USA) to convert the simulated results into predicted 1D \textsuperscript{1}H NMR and nutation spectra.  These spectra are commonly used as measures of \ce{B0} and \ce{B1} homogeneity, and the conversion facilitated comparison with measurements.
The simulated NMR spectra were collected at \SI{90}{\degree} flip-angle and without any shimming compensations.
We have used the material relative permeability values reported by Schenck \cite{Schenck1996} and Wapler \textit{et al.}\cite{Wapler2014}

\subsection{Sample insert geometry}\label{sec:geometry}

The effect of the metal electrodes on the static field \ce{B0}, RF field \ce{B1}, and on resistance-dominated detector noise was investigated. The parameters were processed to deduce all figures of merit, e.g., RF field homogeneity, sensitivity, and spectral resolution of the signal. To this end, several electrode configurations, as depicted in Figure~\ref{fig:setup}b, were considered, and all results were compared to a reference insert containing no electrodes.

The electrode-free reference insert is a simple microfludic channel defined in SU-8 photoresist. The channel height was \SI{90}{\micro \metre} and was sandwiched between two glass wafers, each having a thickness of \SI{210}{\micro \metre}. Starting from this basic microfluidic channel structure, various electrode configurations were chosen with respect to the direction of the \ce{B1} field generated by the Helmholtz pair. From a fabrication perspective, planar disk electrodes in the $zx$-plane were the easiest to  introduce in the microfluidic channel. A working electrode was positioned within the field of view (FOV) of the Helmholtz detector, and the counter electrode was placed \SI{5}{\milli \metre} away within the microfluidic channel. The thickness of the working electrode determined the current induced inside the electrode, and hence also the field perturbation and the dissipated power. For the purpose of this study, three inserts with planar electrode thicknesses of \SI{340}{\nano \metre} ($\sim 10\% \times \delta$), \SI{34}{\nano \metre} ($\sim 1\% \times \delta$), and \SI{3}{\nano \metre} ($\sim 0.1\% \times \delta$) were considered.

Patterning the electrodes on the sidewalls of the microfluidic channel was a second configuration that was explored. It was chosen because it minimises the total electrode surface being penetrated by the \ce{B1} RF field. 
To this end, two high aspect-ratio metallic walls were designed as the active electrodes in the sidewalls of the microfluidic channel, i.e., in the $yz$-plane. The width of the electrodes was set to \SI{30}{\micro \metre} to ensure their mechanical stability, while minimising the cross-section to the \ce{B1} wave front. As shown in Figure~\ref{fig:setup}c, two different situations were considered: placing the sidewall electrodes \SI{2}{\milli \metre} apart (outside the FOV of the Helmholtz coil), or placing inside the FOV of the detector. In the latter case, the optimum distance between the electrodes, w.r.t.\ RF homogeneity, mass sensitivity, and spectral resolution at the sample region, was found to be \SI{1.1}{\milli \metre}.

The third option was to place a structured planar electrode in the FOV of the Helmholtz detector. The three different structures studied were mesh, meander, and comb patterns. The thickness of the electrodes was \SI{34}{\nano \metre}, with a filling factor of \SI{50}{\%}, and a linewidth of \SI{50}{\micro \metre}. These structures offered more resistance to the eddy current and hence less \ce{B1} perturbation, albeit with smaller electrode surface area. Further analysis had been performed through COMSOL simulations.

In terms of its relative magnetic susceptibility to water, copper is a better choice than gold. Nevertheless, gold was chosen as electrode material for the following reasons: i) the value of the standard electrode potential (SEP) of gold is higher than that of copper, i.e., \SI{1.69}{\volt} for gold and \SI{0.339}{\volt} for copper;\cite{Milazzo1978,Bard2017} ii) copper oxidation limits the lifetime of the electrodes especially for the thin films; iii) in case of chitosan deposition, copper ion migration results in metal particles trapped in the gel.\cite{Geng2016}

Of the eight different configurations, four variants were fabricated and experimentally investigated: planar disk electrodes with \SI{35}{\nano \metre} and \SI{3}{\nano \metre} thickness, and both versions of the sidewall electrodes. 

\subsection{Insert microfabrication}

The microfluidic channels were patterned in a layer of SU-8 using UV-lithography on a thin (\SI{210}{\micro \metre}) glass substrate, followed by bonding to another thin glass wafer as a cap layer for sealing. The metallic layers were introduced inside the microfluidic channels either by physical vapour deposition (PVD) and UV-lithography, to form the nano-layers for the electrodes, or by gold electrodeposition to form tracks, counter electrodes, pads, and sidewall electrodes. Complete details of the fabrication processes are given in Section~\ref{SuppSection:fabrication}.

\subsection{Electrodeposition procedure}

\noindent\textit{Sample preparation.} A \SI{1}{\%} w/v chitosan solution was prepared using \SI{10}{\milli \Molar} phosphate buffer solution (PBS) with a \si{pH} of 7.4 in \ce{D2O}. The chitosan flakes were dissolved by adding \ce{DCl} until the \si{pH} of the solution was 3. After stirring overnight, the \si{pH} of the chitosan solution was adjusted to 6 by using \ce{NaOD} dropwise. \SI{50}{\milli\Molar} TSP was added as an NMR reference for the chemical shift at \SI{0}{\ppm}. (All solutions and dilutions use this reservoir buffer to maintain the same \si{pH} 6 and TSP concentration throughout). 

Chitosan was coupled to PEG featuring diacid functionality. For modification of chitosan, a 1:1 CS:PEG molar ratio was chosen in order to maximise the PEG degree of substitution to chitosan amines.\cite{Nordin2019} PEG was activated with EDC and NHS in \SI{20}{\milli \litre} of deionized water for \SI{30}{\minute} at \SI{24}{\celsius}. The PEG:EDC:NHS ratios was 1:0.5:0.5. This solution was then added to a \SI{1}{\%} w/v chitosan solution containing the appropriate quantity of amine reaction centres. The reaction was allowed to proceed for \SI{3}{\hour} at \SI{24}{\celsius}. \ce{NaOD} was added dropwise to the modified polymer solution to form a hydrogel before purification by extensive dialysis.
After the dialysis process, \ce{DCl} was added to redissolve the hydrogel and the \si{pH} was adjusted to \si{pH} 6. The solution was then filtered through a \SI{5}{\micro \metre} syringe filter before use.

\noindent\textit{Electrodeposition.} \textit{In~situ} chitosan electrodeposition was set up using a PID-stabilised current density of \SI{3}{\micro \ampere / \milli \metre \squared}, which was run through the electrode pair for a fixed amount of time to obtain a deposited layer of desired height. 
The deposition procedure is discussed in more detail elsewhere.\cite{Nordin2019} In our case, the current was constantly applied during the entire experiment.  An NMR measurement was made every \SI{14}{min}, which included a \SI{5}{min} delay and \SI{9}{min} of data collection.

\subsection{NMR/MRI data acquisition and analysis}
All magnetic resonance experiments were performed on a \SI{11.74}{\tesla} Avance III Bruker NMR system (Bruker BioSpin, Rheinstetten, Germany) equipped with a Micro5 micro-imaging NMR probe. The detector was a \SI{1.2}{\milli \metre} diameter Helmholtz coil similar to the one reported by Spengler \textit{et al.},\cite{Spengler2016} tuned to \SI{500}MHz (\textsuperscript{1}H Larmor frequency at \SI{11.74}{\tesla}) and matched to \SI{50}{\ohm}. The microfluidic devices containing the different electrode configurations were slid into a slot between the two windings of the Helmholtz pair. The microfluidic channels were designed so that the electrochemical sites are located inside or in close proximity to the FOV of the Helmholtz coil, as shown in Figure~\ref{fig:setup}. The entire setup consisting of the Helmholtz coil with the tuning and matching capacitors and the microfluidic structure was adapted to the commercial probe. This design aligned the middle axis of the microfluidic channel to the $z$-axis of the magnet, which is also the direction of the static magnetic field (\ce{B0}), while the Helmholtz coil generated the excitation RF field along the nominal $y$-axis of the magnet frame of reference (\ce{B1}).

\noindent\textit{Sample preparation.} The sample used for figure-of-merit NMR spectroscopic measurements consisted of \SI{500}{\milli \Molar} sucrose and \SI{50}{\milli \Molar} TSP as the NMR chemical shift reference at \SI{0}{\ppm}, dissolved in deionized water.  The MRI sample was deionized water containing \SI{4}{\milli \Molar} \ce{CuSO4}, and \SI{75}{\milli \Molar} \ce{NaCl}.

\noindent\textit{NMR spectroscopy.} \textsuperscript{1}H NMR experiments were performed using TopSpin~3.5, the operating and processing software for Bruker NMR spectrometers. A one-dimensional NMR experiment was carried out on the sucrose sample at \SI{30}{\celsius}. The applied power was \SI{0.8}{\watt} and shimming was performed manually up to the second order parameters. Each spectrum represents \SI{64}{scans}. 18000 data points were collected for each scan over a spectral width of \SI{20}{\ppm}, the relaxation delay being set to \SI{5}{\second}. The resulting signal was Fourier transformed after multiplication with an exponential function equivalent to \SI{0.3}{\hertz} line broadening. The full width at half maximum of the \ce{TSP} peak was calculated as a measure for the spectral resolution.

A nutation experiment was performed in order to determine the \ce{B1} field homogeneity. The nutation spectrum consists of 300 single scans using \SI{0.8}{\watt} applied power with an increment of \SI{1}{\micro \second}. A relaxation delay of \SI{5}{\second} was set between two consecutive scans. Each data point in the nutation spectrum, represented by a single NMR peak, was integrated and plotted as part of a continuous curve to facilitate comparison with the simulated experiments. 

\noindent\textit{MRI.} \ce{B1} distributions for inserts with different electrode configurations were studied through MRI experiments (ParaVision~6.0.1) using a sucrose sample. The experimental parameters of the \textit{Flash} sequence were set as follows: TR/TE - \SI{200}{\milli \second}/\SI{4}{\milli \second}, slice thickness -  \SI{100}{\micro \metre} along the $y$-axis, in-plane resolution - $40\times 40 $ \SI{}{\squared \micro \metre}, Tscan - \SI{53}{\minute}, FOV $2\times 2$ \SI{}{ \milli \metre \squared}, bandwidth - \SI{10}{\kilo \hertz}, and 512 averages. The duration of the excitation pulse was fixed at \SI{1.1}{\milli \second} and its power level was adjusted to achieve different flip angles.

\textit{FieldMap} sequences were performed to study \ce{B0} distribution experimentally, at the detection zone. The \textit{FieldMap} sequence applies a 3D double gradient-echo image sequence with two different echo times, and based on the phase and magnitude differences between the echos, it calculates the \ce{B0} distribution.\cite{Gruschke2012} The experimental parameters of the \textit{FieldMap} sequence were set as follows: TR - \SI{35}{\milli \second}, TE1/TE2 - \SI{1.31}{\milli \second}/\SI{7.03}{\milli \second}, isotropic voxel size - \SI{90}{\micro \metre}, Tscan - \SI{76}{\minute}, FOV $2.9\times 2.9\times 2.9$ \SI{}{\cubic \milli \metre}, bandwidth - \SI{30}{\kilo \hertz}, number of averages - \SI{128}{}, flip angle - \SI{50}{\degree}. SNR threshold was set to \SI{5}{} for data collection to minimise noise-induced image distortions. The slice which corresponds to the sample position in the $zx$-plane has been considered for representation and analysis. The standard deviation of the collected map was calculated and studied as a measure for the \ce{B0} homogeneity. For this purpose, the outermost pixels of the acquired image were filtered out to further reduce the noise contribution.

\noindent\textit{In~situ electrochemistry.} A reference one-dimensional NMR experiment was initially performed on the sample containing only the CS solution at \SI{30}{\celsius}, to identify the chitosan signal at $\sim$\SI{3.3}{\ppm} (singlet, \textsuperscript{1}H). Electrodeposition was monitored by collecting a series of 1D \textsuperscript{1}H NMR spectra as a function of time. 
Each NMR spectrum was a result of \SI{200}{scans}, each containing 34000 data-points over a spectral width of \SI{20}{\ppm}, the relaxation delay being set to \SI{1}{\second}. The resulting signals were multiplied by an exponential function equivalent to \SI{0.3}{\hertz} line broadening prior to Fourier transform. 

%% file: 03_Results.tex
\section{Results and discussions}

In order to enable \textit{in-situ} NMR monitoring of electrochemical processes in a lab-on-a-chip environment, metal electrodes must be introduced inside the microfluidic channel, ideally in the NMR sensitive volume. Two effects must therefore be managed so that high resolution NMR spectra with minimal sensitivity loss are still obtainable: the influence on the homogeneity of the primary static magnetic field \ce{B0}, and the performance of the excitation magnetic field \ce{B1}.  Magnetic susceptibility mismatches between materials will broaden NMR signals in the region of material interfaces perpendicular to \ce{B0}, degrading overall spectral resolution and signal-to-noise ratio (SNR).  Conductive structures perpendicular to the \ce{B1} field front will carry induced eddy currents, depending on the cross-section of the structure facing the field front and the resistance of the current path. The net effect is the conductive material effectively shielding the sample from \ce{B1}, therefore reducing sensitivity.

Taking the gold skin depth ($\delta\sim$\SI{3.37}{\micro \metre} at \SI{500}{\mega \hertz}) into account, eight different electrode topologies (previously introduced in Section~\ref{sec:geometry}) were simulated, critically evaluated, and compared to the situation when no electrode is inserted in the NMR sensitive volume. Four such topologies were fabricated and experimentally tested. All results are presented in a condensed manner in Table~\ref{tab:FigOfMer}.

\input{table.tex}

\subsection{Performance analysis of simulated electrode topologies}

\subsubsection{\ce{B1} distribution}

\begin{figure*}[htpb]
    \centering
    \includegraphics[width=0.8\linewidth]{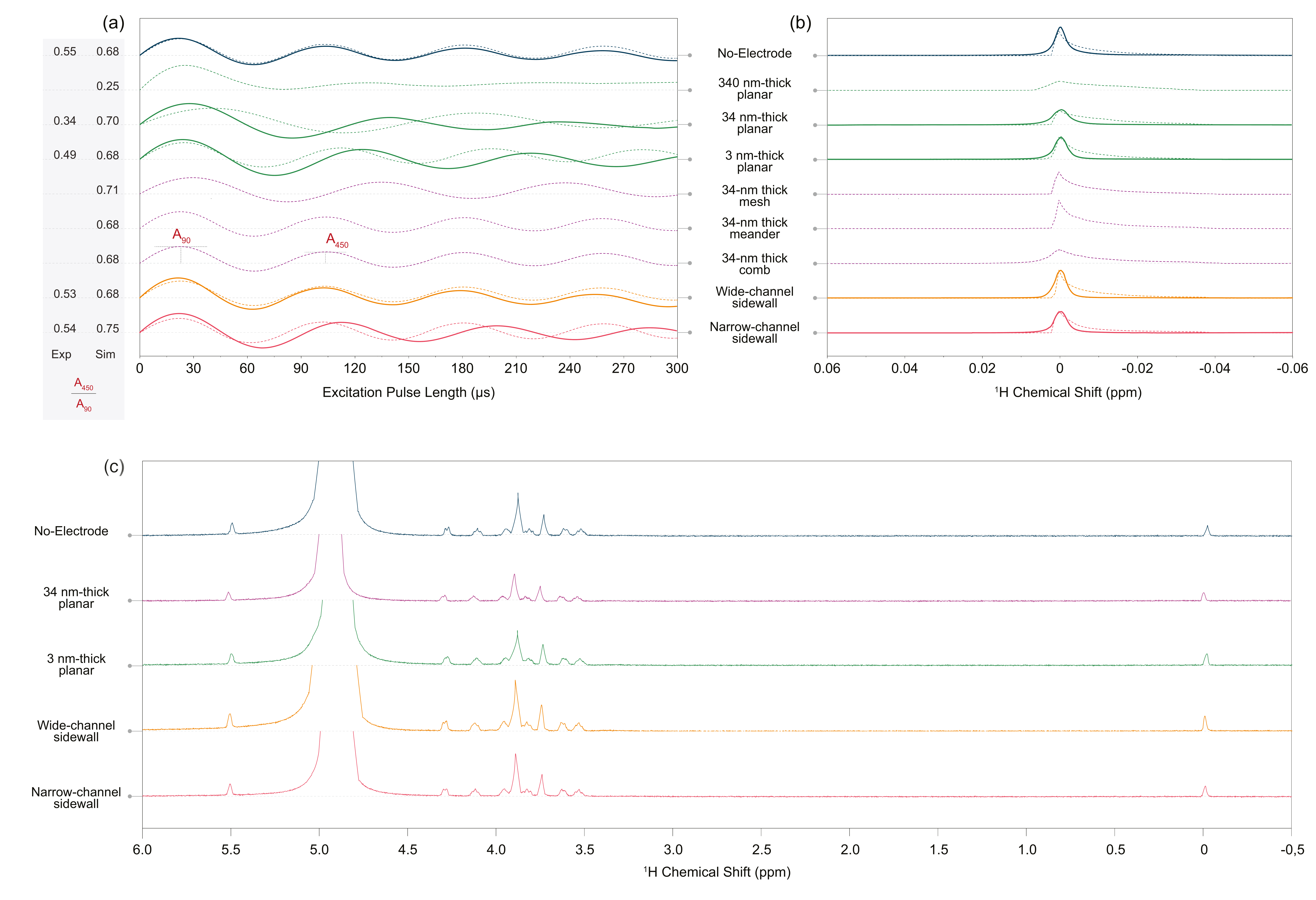}
    \caption{(a) Simulated (dash) and measured (solid)  nutation curves for different inserts: electrode-free, planar disk electrodes (\SI{340}{\nano \metre}-, \SI{34}{\nano \metre}-, and \SI{3}{\nano \metre}- thickness), structured planar (\SI{34}{\nano \metre} thickness) electrodes (mesh, meander, and comb), sidewall electrodes (wide channel (\SI{2}{\milli \metre}), narrow channel (\SI{1.1}{\milli \metre})).
    (b) Simulated (dash) and measured (solid) \textsuperscript{1}H NMR spectra of the TSP signal in different configurations. 
    (c) Measured \textsuperscript{1}H NMR spectra of the sucrose sample in five different configurations. 
 }
    \label{fig:NMR_results}
\end{figure*}

The nutation curve reflects the progression of the magnetisation as a function of the excitation pulse length (at a fixed applied power) and can be used to estimate the excitation field performance. The damping of the curve reports on the \ce{B1} excitation field homogeneity: a fast decay is the result of poor homogeneity. The frequency of the nutation curve represents the conversion efficiency of the applied electrical power to the \ce{B1} field at the sample region. For a given power, a higher sensitivity of the coil or a lower power dissipation in the system leads to a higher nutation frequency. The normalised simulated nutation curves for different inserts are shown in Figure~\ref{fig:NMR_results}a. The ratio between the signal amplitude at the pulse length corresponding to flip angles of \SI{450}{\degree} and \SI{90}{\degree} - \textbf{$A_{450}/A_{90}$} - is considered as a figure of merit for \ce{B1} homogeneity and is used to compare different insert topologies.

From the point of view of \ce{B1} performance, a first conclusion offered by the comparative presentation in Table~\ref{tab:FigOfMer} and Figure~\ref{fig:NMR_results}a was that almost all structures perform in a similar manner: their simulated \ce{B1} field homogeneity was at least as good as for the electrode-free case, with the factor $A_{450}/A_{90}$ between 68\% and 75\%.
The only exception was the \SI{340}{\nano \metre}-thick planar electrode that degraded the \ce{B1} field homogeneity to $A_{450}/A_{90}$=25\%, i.e., almost one third compared to the electrode-free case. Three structures (\SI{34}{\nano \metre}-thick disk, the mesh, and the narrow channel sidewall electrode) showed slightly improved field homogeneity.

The eddy currents induced in the disk metallic structures depleted the \ce{B1} field from the centre and concentrated it around the edges of the electrodes. This field distortion increased with the thickness of the metal layer. For a \SI{34}{\nano \metre}-thick layer, this effect led to a flattened field distribution, which enhanced the \ce{B1} field homogeneity. The mesh electrode provided multiple closed loops for the eddy current paths, by spreading the field over the entire detection zone, and further enhancing the \ce{B1} homogeneity to $A_{450}/A_{90}$=71\%. 
The inter-electrode distance for the narrow channel sidewall configuration was optimised for \ce{B1} field homogeneity, reaching a value of $A_{450}/A_{90}$=75\%, achieved by both constraining the sample inside the certain region which had the best field homogeneity, and compressing the field inside the sample volume using the sidewall electrodes.

\subsubsection{Sensitivity}

According to the reciprocity law, for a uniformly distributed and polarised sample, the total area under an NMR peak is a measure of the average \ce{B1} and is independent of the spectral resolution, which is influenced by the homogeneity of the static \ce{B0} field.
In Figure~\ref{fig:NMR_results}b), the signal amplitudes were scaled so that they all had the same noise level. The mass sensitivity, defined as the average \ce{B1} per collected noise amplitude for a unit sample volume,\cite{Webb2013} was studied in order to compare the performance of different insert topologies. Therefore, the area under the normalised NMR spectrum collected for each electrode geometry, relative to the electrode-free case - \textbf{$S_{rel}$} - was considered as a second figure of merit.

As expected from the analytical expressions given by Chen \textit{et al.},\cite{Chen2017} the thickest planar disk  electrode showed the worst performance in terms of mass sensitivity - $S_{rel}$= 86.4\%. For all other structures except for the narrow channel sidewall configuration, the penalty to be paid in comparison with the electrode-free configuration, in terms of mass sensitivity, was $<5\%$. The narrow channel sidewall electrode configuration clearly outperformed all other designs. 
The sensitivity enhancement for the narrow channel sidewall structure could also be observed through the nutation frequency (Figure~\ref{fig:NMR_results}a). The slightly higher ($1.04 \times$) nutation frequency of the narrow channel sidewall electrode configuration, compared to the electrode-free insert, suggests that the presence of the vertical electrode is constructive and enhances the overall sensitivity.

\subsubsection{\ce{B0} distribution}

The line-shape of a well-resolved NMR signal is considered a measure of \ce{B0} distortion, with signals deviating from an ideal Lorentzian reflecting poor \ce{B0} homogeneity. Figure~\ref{fig:NMR_results}b presents the simulated NMR signal for the eight different electrode configurations, as well as the electrode-free structure for comparison. Simulations only account for the Helmholtz coil and the insert materials (as introduced in Section~\ref{sec:geometry}), without considering any additional elements of the probe/circuit. A shoulder is observed next to the signal for all structures including the case of the electrode-free insert, and is a result of the susceptibility mismatch between the sample, glass substrates, and air. 

The planar electrode configurations generate a material interface in the immediate vicinity of the FOV, which degrades the spectral resolution. As expected, this degradation was found to be more severe with an increase of the thickness of the metal layer ( $\textit{FWHM} = \SI{16.59}{ppb}$ for \SI{340}{\nano \metre} and $\textit{FWHM} = \SI{9.05}{ppb}$ for \SI{34}{\nano \metre}). However, for a thickness of \SI{3}{\nano \metre}, the spectral resolution returned to a value comparable to the electrode-free configuration, and similar to other structures which had been evaluated. It is interesting to note that all patterned planar electrodes improved the spectral resolution in comparison to a disk electrode with the same thickness, the \textit{FWHM} value being comparable to the electrode-free case, as shown in Table~\ref{tab:FigOfMer}. The spectra collected from the wide and narrow channel sidewall configurations have \textit{FWHM} of \SI{4.18}{ppb} and \SI{4.57}{ppb}, which were similar to the \textit{FWHM} of the spectrum from the electrode-free insert (\SI{4.28}{ppb}). This was achieved by extending the electrodes along \ce{B0} to shift the material interface far from the FOV. 

\subsubsection{Choice of topologies to be fabricated}

The simulation results reveal two clear extremes. The disk electrode structure with a thickness of \SI{340}{\nano \metre} showed the worst figures of merit for all the characteristics that had been investigated: \ce{B1} uniformity, mass sensitivity, and \ce{B0} uniformity. 
On the other hand, the narrow channel sidewall electrode outperformed all other structures, including the electrode-free configuration, by exhibiting superior $A_{450}/A_{90}$ factor and mass sensitivity, while the spectral resolution was only slightly less than the electrode-free configuration. Therefore, we excluded the \SI{340}{\nano \metre} planar electrode structure from fabrication and further experimental investigation, but retained the narrow channel sidewall electrode.

Three additional structures were selected for fabrication and experimental characterisation.  These geometries had similar simulated performance and were, importantly, compatible with the electrochemistry experiment designed in Section~\ref{sec:CS_ED}: the wide channel sidewall electrode structure and, for the ease of fabrication, the \SI{34}{\nano \metre} and \SI{3}{\nano \metre} disk electrode structures. These four structures together with the electrode-free configuration were investigated experimentally and are discussed in the next section.

\subsection{Performance analysis of fabricated electrode topologies}

\subsubsection{\ce{B1} distribution}

Tuning and matching of the NMR probe affects the noise level, power transmission efficiency, and nutation frequency of experimental measurements. However, the decay rate of the nutation signal is independent of tuning and matching. Therefore, the measured nutation spectra are good and independent indicators with which to determine the homogeneity of the \ce{B1} field for the five different fabricated inserts.

The absolute values for the $A_{450}/A_{90}$ factor, extracted from the nutation curves in Figure~\ref{fig:NMR_results}a and presented in Table~\ref{tab:FigOfMer}, were found to be smaller compared to the simulated values. This is attributed to the fabrication tolerances for the Helmholtz detector, e.g., Helmholtz coil size, slight misalignment of the windings, the distance between two windings, and the effect of the extra tracks which have not been considered in the simulations, as well as the alignment of the insert with respect to the coil.

In contrast with the simulated results, the experiment showed an inferior performance of the structures with planar electrodes in terms of \ce{B1} field uniformity when compared to the electrode-free configuration. We attribute this to the fact that the planar electrode structure introduces severe \ce{B0} distortions (as discussed in Section~\ref{sec:B0_exp}) which cannot be fully compensated by shimming. This broadens the linewidth and hence reduces the height of the peak, the effect being more pronounced for higher flip angles due to the additional signal phase deformation.

The two sidewall electrode configurations confirmed experimentally the simulated behaviour with a \ce{B1} field homogeneity in the same range as the electrode-free configuration: wide channel sidewall electrode: $A_{450}/A_{90}$=53\% and narrow channel sidewall electrode: $A_{450}/A_{90}$=54\%, versus electrode-free: $A_{450}/A_{90}$=55\%. 

\begin{figure}
    \centering
    \includegraphics[width=\linewidth]{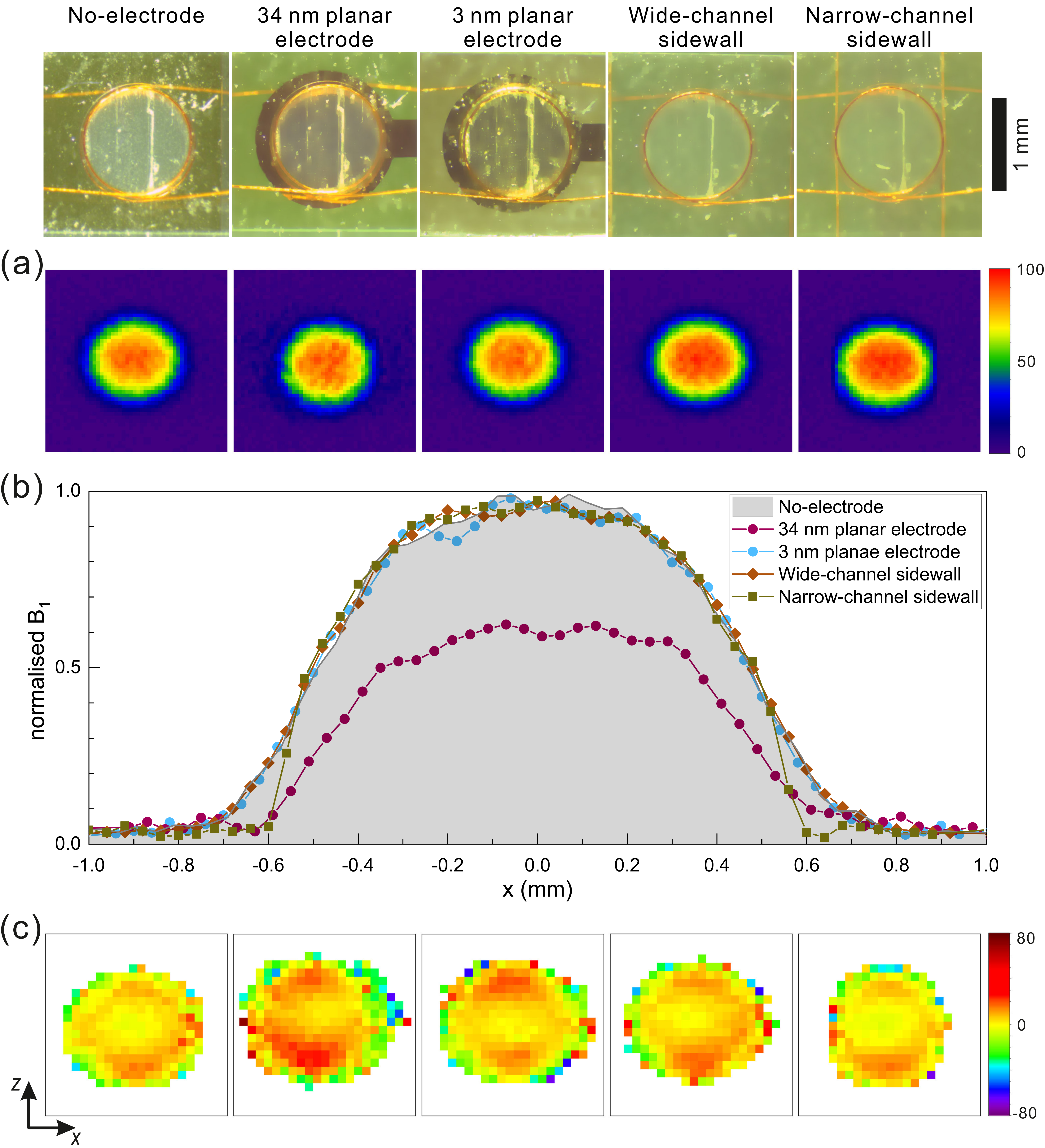}
    \caption{(a) Normalised distribution of the \ce{B1} field (\%) at the sample volume excited with \SI{19.2}{\micro \watt} power for different types of inserts. \ce{B1} distributions using different excitation powers can be found in Figure~\ref{fig_sup:B1map_MRI}. (b) Normalised profile of \ce{B1} along the x-axis at excitation power of \SI{19.2}{\micro \watt}.  (c) \ce{B0} field map at the detection zone of the coil in ppb.}
    \label{fig:B1map_MRI}
\end{figure}

MRI \ce{B1} maps at a flip angle corresponding to \SI{19.2}{\micro \watt} for all five configurations, and verified experimentally, are presented in Figure~\ref{fig:B1map_MRI}a and confirm the lower homogeneity of the planar electrodes (the \ce{B1} profiles at various excitation powers are presented in Figure~\ref{fig_sup:B1map_MRI}). In order to highlight the field homogeneity, the maps were scaled to unity. 
Figure~\ref{fig:B1map_MRI}b depicts the profile of the \ce{B1} field along the $x$-axis at the middle of the detection zone. For the narrow channel insert, the measurements confirmed that the field homogeneity was enhanced and that a steep drop of the field happened as the sample volume was restricted by the metallic sidewall electrodes.

\subsubsection{Sensitivity}

Figure~\ref{fig:NMR_results}b, c presents NMR spectra of the sucrose sample collected using the different inserts. Similarly, the relative mass sensitivity of each configuration was calculated by considering the integral of the TSP signal (located at \SI{0}{\ppm}) normalised by that of electrode-free geometry as the reference - $S_{TSP}$. The results are presented in Table~\ref{tab:FigOfMer}. The difference between the absolute values of the simulated results and measurements is attributed to different tuning/matching conditions, hence signal transfer efficiency and noise performance. However, the general trend is reproduced and these results confirm the superior performance of the narrow channel sidewall configuration.

\subsubsection{\ce{B0} distribution}\label{sec:B0_exp}

Spectral resolution of experimental NMR spectra strongly depends on the ability to correct \ce{B0} imperfections by using shim fields. The NMR signal of the sample for different configurations, after the optimal shim settings, were identified and are presented in Figure~\ref{fig:NMR_results}b, c. The \textit{FWHM} value for TSP peak was considered as the measure for the spectral resolution.
Similar to the simulation results, the \SI{34}{\micro \metre}-thick planar structure shows the widest linewidth, whereas the \SI{3}{\micro \metre}-thick planar, wide-channel sidewall, and narrow-channel sidewall electrode structures similarly perturb \ce{B0}. The \textit{FWHM} for the electrode-free insert has a slightly better spectral resolution. All \textit{FWHM} values are reported in Table~\ref{tab:FigOfMer}.

In order to investigate the \ce{B0} distribution more precisely and avoid shimming influence on the results, the \ce{B0} maps of different configurations were collected and are plotted in Figure~\ref{fig:NMR_results}b. The standard deviation calculated from the \ce{B0} map was used as a second figure of merit for \ce{B0} distribution analysis. Theses values are presented in Table~\ref{tab:FigOfMer}. In all configurations, the detection volume was surrounded by a rim whose voxels were severely distorted due to the the partial volume effect (the sample does not fully occupy a voxel) and low \ce{B1} field.  To calculate the standard deviation of the colour maps, a rim of one pixel was excluded to minimise the noise contribution. At the bottom right side of the detection zone, similar field distortions were observed in all the \ce{B0} field maps. This distortion originated from the solder which connected the chip to the PCB (see Figure~\ref{fig:setup}).

The \ce{B0} pattern at the sensitive zone of the coil showed that the planar \SI{34}{\nano \metre} electrode distorted the static field especially at the top and bottom edges of the electrode, which perfectly aligned the material interface intersections with \ce{B0}. The ultra-thin electrodes (\SI{3}{\nano \metre} thickness) introduced less perturbations; however, the overall pattern appeared similar as expected. Conversely, the narrow channel sidewall electrodes had almost no effect on the overall field pattern, except for the left and right edges where the \textsuperscript{1}H NMR signal was excluded due to the presence of the electrodes. These results correlated with the measured spectral \textit{FWHM} results.

\subsection{Fabrication challenges and opportunities}
According to the measurements and FEM simulations, the sidewall electrodes, with major surfaces parallel to both \ce{B0} and \ce{B1}, and ultra-thin (less that \SI{0.1}{\%} of skin-depth) planar electrodes were found to have the best figure-of-merit values. 
The challenge of introducing conductive structures for \textit{in~situ} electrochemical analyses with high-field NMR is reflected by there being only a few reports available. \cite{da_Silva_Electrochemical_2019, Gomes_Strong_2017, Ni_A_2017}  Micro-fabrication introduces new opportunities. On one hand, ultra-thin metallic layers, which are NMR transparent, can be achieved by rather standard MEMS techniques. On the other hand, microfabrication enables the construction of precisely oriented high aspect-ratio electrodes, which have a minimum footprint to eliminate any \ce{B1} and \ce{B0} distortions. Here, we have demonstrated a fabrication technique based on UV-lithography and gold-electroplating to manufacture such electrodes with an aspect ratio (height to width) of $\sim 3$. A further reduction of the width of the electrodes requires more sophisticated lithography techniques, e.g., 2PP, X-Ray or e-beam lithography. As an alternative, the distance between the electrodes was optimised in this work.

Since the narrow channel sidewall electrode configuration proves to be the best compromise in terms of \ce{B1} field homogeneity and mass sensitivity, as well as in terms of \ce{B0} field distortion, i.e., spectral resolution, this structure was further used for the experiment of NMR \textit{in~situ} monitoring of chitosan electrodeposition.

\subsection{\textit{In~situ} electrodeposition of chitosan}\label{sec:CS_ED}
Standard electrochemical analyses use three-electrode setups consisting of working, counter, and reference electrodes.  Nevertheless, application of a potential between a single pair of electrodes can still be exploited for a variety of uses. 

Using this electrode configuration, the compatibility of an electrochemical experiment with NMR spectroscopy within a lab-on-a-chip environment was demonstrated using chitosan electrodeposition. This experiment is attractive first because of its relative simplicity, requiring only a voltage applied between 2 electrodes to initiate water hydrolysis, resulting in a local pH gradient as required for chitosan hydrogel deposition. Second, the NMR spectrum of chitosan in solution is significantly different compared to the gel state, and thus the deposition process, as a function of time, can be monitored. Third, chitosan can be chemically modified so that NMR signals from covalently attached molecules can also be monitored as a function of gel formation. Finally, the chitosan hydrogel architecture can be controlled\cite{Nordin2019}.

\begin{figure}[!ht]
    \centering
    \includegraphics[width=\linewidth]{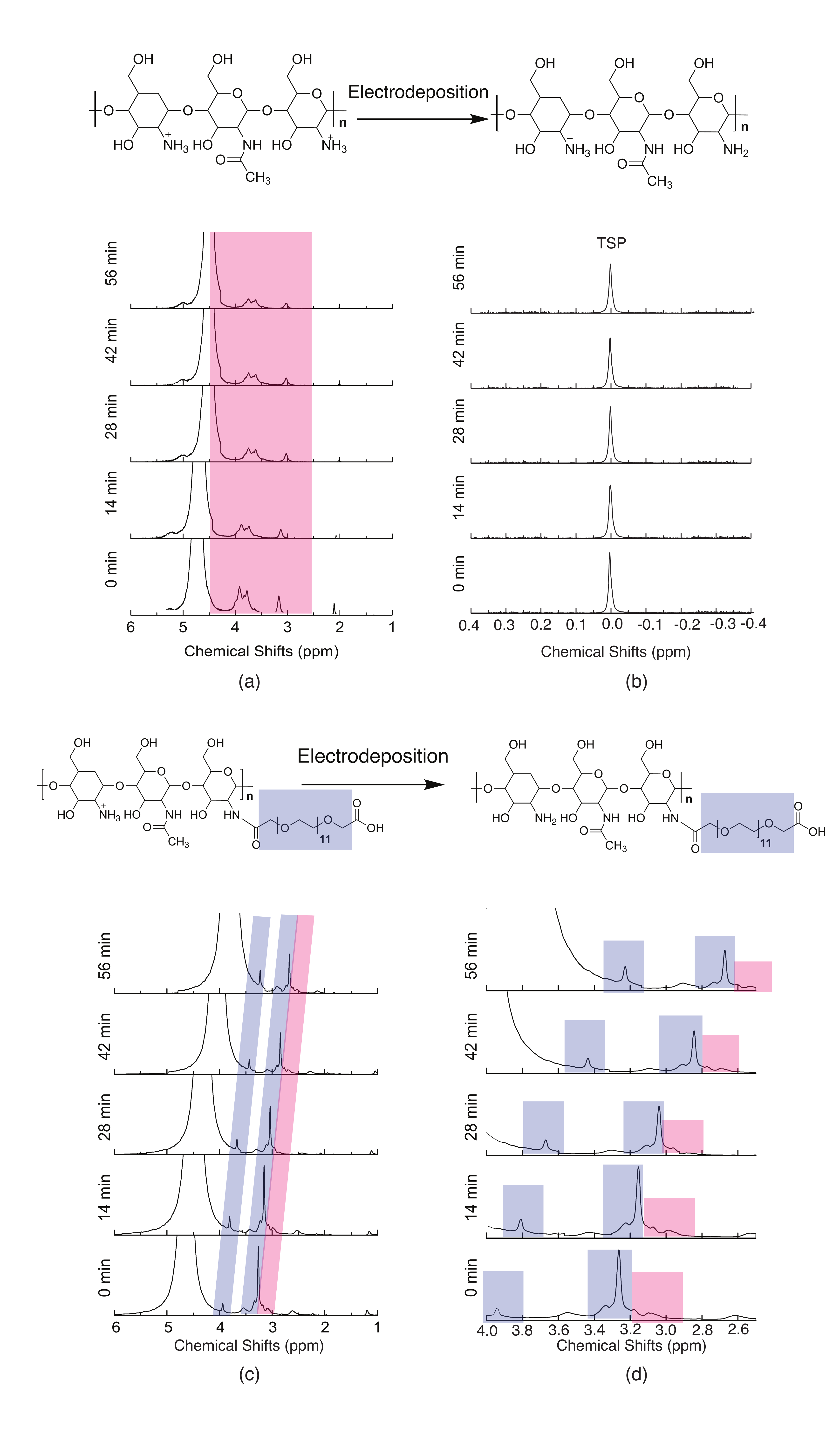}
    \caption{\textit{In~situ} $^1$H NMR monitoring of chitosan (CS) electrodeposition. Results for unmodified CS. (a) CS chemical shift region (pink highlight). (b) TSP signal 0 ppm. Results for CS-PEG: (c) CS-PEG signals (PEG signals in blue highlight). (d) Zoom of 2.6-4.0 ppm region to highlight the PEG signals.  The time between each spectrum was determined by a \SI{5}{\minute} delay + \SI{9}{\minute} NMR acquisition. In both cases, the water resonance was observed to shift due to a change in pH. The current used to drive electrodeposition was applied during the entire experiment. 
    }
    \label{fig:spectra}
\end{figure}

To perform the \textit{in~situ} experiment, a chitosan (CS) solution was injected into the narrow-channel sidewall insert (1\% CS, \SI{50}{\milli \Molar} TSP, \si{pH} 5.5).  Two versions of chitosan were investigated: the native biopolymer, and one chemically modified with polyethyleneglycol (PEG). Prior to applying a current, a \textsuperscript{1}H NMR spectrum of the solution was measured and NMR calibrations were performed.
After the reference spectrum was measured, a current was applied (\SI{3}{\micro \ampere / \milli \metre \square} over a working electrode area of \SI{0.42}{\square \milli \metre}).  The current flowed for the entire experiment, and after intervals of \SI{5}{\minute} a new \textsuperscript{1}H NMR spectrum was acquired. This cycle was repeated so that a total of 5 spectra were measured, the results are summarised in Figure~\ref{fig:spectra}.

The \textsuperscript{1}H NMR signals from chitosan in solution are clearly visible in the chemical shift range 3 - 4 ppm (pink region in Figure~\ref{fig:spectra}a, b).  As expected, these signals begin broadening as a function of deposition time.  The broadening results from constrained molecular motion as would be expected in the hydrogel state.  A similar result was observed for chitosan when modified with PEG (Figure~\ref{fig:spectra}c, d).  In contrast, the PEG signal (\SI{\sim 3.8}{\ppm} and \SI{\sim 3.3}{\ppm}, blue region)  remained relatively sharp as the hydrogel was formed, suggesting this highly hydrated polymer maintained a degree of molecular motion. This was an interesting result, since bi-functional PEG can be used to attach interesting molecules to chitosan, which would then be potentially decoupled from the deleterious line-broadening effects when deposited as the hydrogel. This feature is currently being explored in continuing work in our group.

%% file: table.tex
\begin{table*}[pb]
 \small 
  \begin{center}
    \caption{Summary of the NMR/MRI figures of merit for performance evaluation of different inserts. The colour code highlights the comparison to the no-electrode case.    }
    \label{tab:FigOfMer}
\begin{tabular}{p{3.5cm} p{1.6cm} p{1cm} p{2.2cm}  p{0.01cm} p{1.6cm} p{1.1cm} p{2.2cm} p{0.8cm}} \toprule
    \multirow{2}{*}{Configurations} 
        & \multicolumn{3}{c} {\textbf{Simulation}} & & \multicolumn{4}{c} {\textbf{Experimental}} \\ 
        \cmidrule{2-4} \cmidrule{6-9}
        & $A_{450}/A_{90}$ (\%) & $S_{rel}$ (\%) & \textit{FWHM} (ppb/Hz) & & $A_{450}/A_{90}$ (\%) & $S_{TSP}$ (\%) & \textit{FWHM} (ppb/Hz) & $\sigma_{B0}$  \\ \midrule
    
    \textcolor{RoyalBlue}{Electrode-free} & \textcolor{RoyalBlue}{68}  & \textcolor{RoyalBlue}{100}  & \textcolor{RoyalBlue}{4.28 / 2.14} & & \textcolor{RoyalBlue}{55}  & \textcolor{RoyalBlue}{100}  & \textcolor{RoyalBlue}{9.5 / 4.8} & \textcolor{RoyalBlue}{3.1}  \\ 
    
    \textcolor{red}{\SI{340}{\nano \meter}-thick disk} & \textcolor{red}{25}  & \textcolor{red}{86.4}  & \textcolor{red}{16.59 / 8.30} & & - & - & - & - \\ 
    
    \textcolor{red}{\SI{34}{\nano \meter}-thick disk} & \textcolor{gray}{70}  & \textcolor{gray}{94.9}  & \textcolor{red}{9.05 / 4.51} & & \textcolor{red}{34}  & \textcolor{red}{85}  & \textcolor{red}{13.3 / 6.7} & \textcolor{red}{5.3}  \\ 
    
    \SI{3}{\nano \meter}-thick disk & \textcolor{gray}{68}  & \textcolor{gray}{99.4}  & \textcolor{gray}{4.62 / 2.31} & & \textcolor{gray}{49}  & \textcolor{gray}{93}  & \textcolor{gray}{11.2 / 5.6} & \textcolor{gray}{3.7} \\ 
    
    \SI{34}{\nano \meter}-thick mesh & \textcolor{gray}{71}  & \textcolor{gray}{97.7}  & \textcolor{gray}{4.42 / 2.21} & & - & - & - \\ 
    
    \SI{34}{\nano \meter}-thick meander & \textcolor{gray}{68}  & \textcolor{gray}{98.9}  & \textcolor{gray}{4.54 / 2.27} & & - & - & - \\ 
    
    \SI{34}{\nano \meter}-thick comb & \textcolor{gray}{68}  & \textcolor{gray}{100}  & \textcolor{gray}{5.35 / 2.68} & & - & - & - \\ 
    
    Wide-channel sidewall & \textcolor{gray}{68}  & \textcolor{gray}{99.0}  & \textcolor{gray}{4.18/ 2.09} & & \textcolor{gray}{53}  & \textcolor{gray}{100}  & \textcolor{gray}{11.4 / 5.7} &  \textcolor{gray}{3.6} \\ 
    
    \textcolor{PineGreen}{Narrow-channel sidewall} & \textcolor{PineGreen}{75}  & \textcolor{PineGreen}{154.8} & \textcolor{gray}{4.57 / 2.29} & & \textcolor{gray}{54}  & \textcolor{PineGreen}{141} & \textcolor{gray}{11.7 / 5.9} & \textcolor{gray}{3.5} \\ 
    
\bottomrule

\end{tabular}
 \end{center}
\end{table*}

%% file: 04_Conclusions.tex
\section{Conclusions}

Introducing metallic electrodes into the NMR detection zone is a challenge to be addressed when performing \textit{in~situ} NMR characterisation of electrochemical processes. The main obstacles to overcome are SNR degradation caused by \ce{B1} inhomogeneity and power dissipation in the electrodes, as well as line broadening caused by magnetic susceptibility mismatches. The requirements become even more stringent when integration in a lab-on-a-chip system is required. In order to propose solutions to circumvent these issues, we simulated eight different configurations. Four of those configurations were fabricated and further analysed. The configurations including sidewall electrodes, parallel to both \ce{B0} and \ce{B1}, and the configuration with ultra-thin (less that \SI{0.1}{\%} of skin-depth) planar electrodes are found to be the best choices according to the FEM simulations, as well as to the NMR and MRI measurements. Chitosan electrodeposition is presented as an exemplary electrochemistry experiment to be observed by NMR. The chitosan gel state, localised to the electrode surface, is exploited for analyte immobilisation either chemically or physically within the NMR detection volume.  In this direction, the quantification of \textit{in~situ} electrodeposition of PEG-modified chitosan was carried out.

We conjuncture that the results of this work will be highly relevant for other applications and processes that involve the presence of metallic structures, when these are combined with NMR monitoring. Electrokinetic separation methods, where analytes move through electrolytes under the influence of an applied electric field, are widely used in $\mu$-TAS environments and require the presence of metal electrodes within the device. Dielectrophoresis is another phenomenon extensively used in lab-on-a-chip system to immobilise dielectric particles in non-uniform electric fields. More recently, droplet microfluidics which involves manipulation of droplets of analyte using a planar array of electrodes has been combined with \textit{in~situ} NMR monitoring facing similar challenges.

The methodology presented here can also be extended to other electrode materials compatible with microfabrication and relevant to various electrochemical studies. For example, electrode materials including Ag/AgCl, platinum, and carbon are rather common in 3-electrode electrochemical setups, and indium-tin-oxide (ITO) can be used as transparent electrode, thus enabling direct optical observation at the surface of the electrode where deposition is taking place, or under the electrode in the process chamber.

All similar applications of interest to the lab-on-a-chip community will benefit from the present study, opening new avenues by hyphenating a very chemically specific characterisation method such as NMR with electrochemistry.

%% file: 05_Acknowledgements.tex
\section{Acknowledgements}

 H.D, N.N, N.M, and V.B. acknowledge funding from the Deutsche Forschungsgemeinschaft for the project Bio-PRICE [DFG BA 4275/4-1, DFG MA 6653/1-1], L.B. acknowledges partial support from the Carl Zeiss Stiftung. L.B. and J.K. acknowledge financial support from the the European Union [Grant H2020-FETOPEN-1-2016-2017-737043-TISuMR]. N.N wishes to thank the University of Malaya and Ministry of Higher Education of Malaysia for the scholarship (SLAB Scheme) and the BioInterfaces International Graduate School (BIF-IGS, www.bif-igs.kit.edu).  The authors acknowledge support by the Karlsruhe Nano Micro Facility (KNMF), a Helmholtz Research Infrastructure at Karlsruhe Institute of Technology.

%% file: 06_ElectronicSupplementaryInfo.tex
\beginsupplement

\subsection{Micro-fabrication process of the inserts} \label{SuppSection:fabrication}

As presented in Section~\ref{sec:geometry}, different electrode configurations were introduced starting from the basic microfluidic channel structure.
The microfluidic channel was fabricated on a 4-inch MEMpax substrate by spincoating \SI{90}{\micro \metre} SU-8 followed by UV-lithography (EVG\textregistered620 EV~Group) to define the channel structure. \SI{5}{\micro \metre} SU-8, being spincoated on the second MEMpax wafer, served as the adhesion layer between the two MEMpax wafers which were further bonded using a compression bonding machine (EVG\textregistered510 EV~Group). 
The assembly was subjected to UV exposure (\SI{300}{\milli \joule \per \centi \metre \squared}) and baked (\SI{4}{\hour} at \SI{95}{\celsius}) to enhance the mechanical and chemical stability of the bond. The inlet and outlet of the microfluidic channel were drilled on the top glass wafer using a nanosecond laser (PIRANHA\textregistered \ ACSYS). The assembly was diced into the individual chips using the same laser. 

\begin{figure}[ht]
    \centering
    \includegraphics[width=0.45\linewidth]{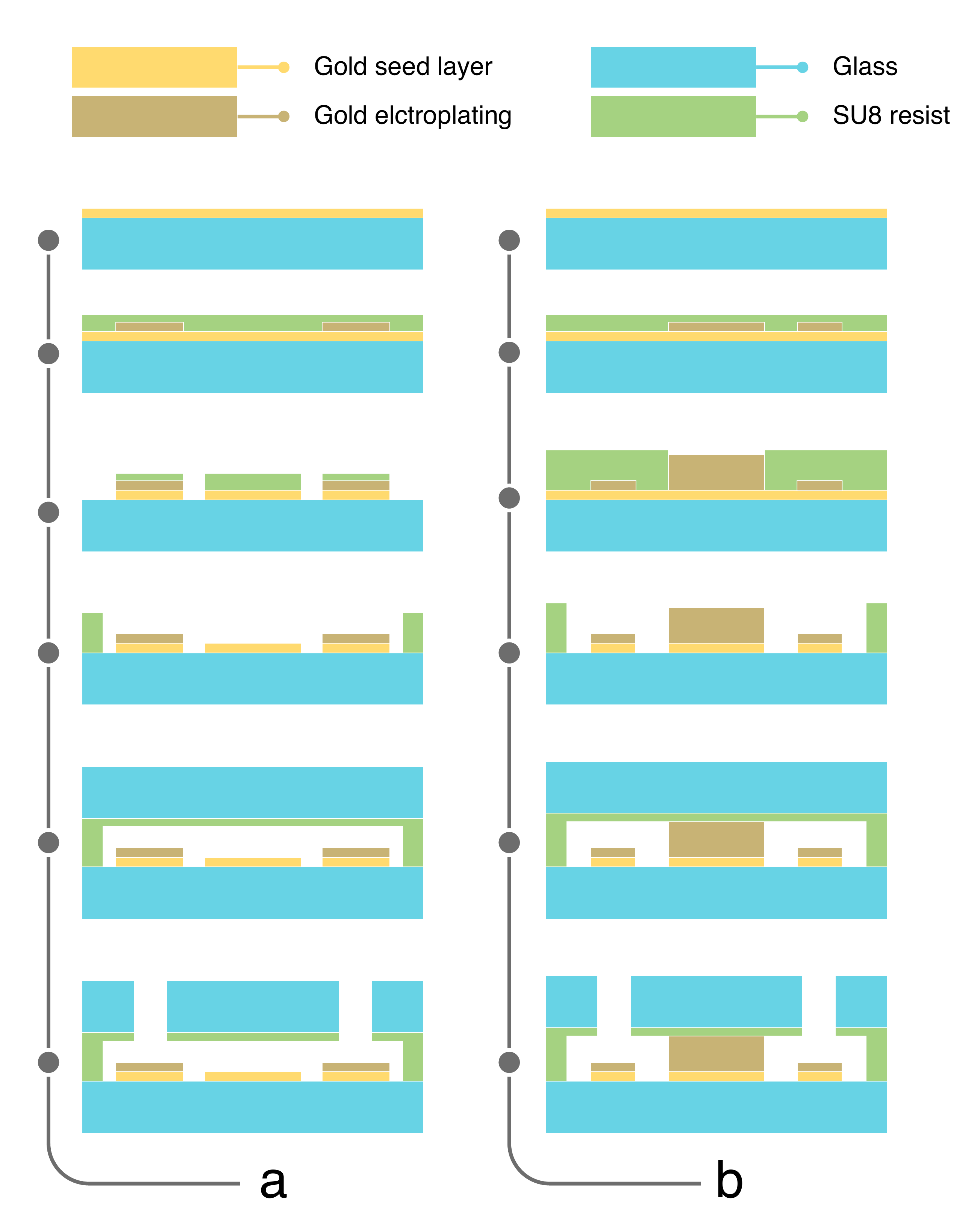}
    \caption{(a) The inserts with planar electrodes are fabricated by electroplating the counter electrode on the seed layer, developing active electrode through the seed layer, patterning the channel, and drilling in-/oulets. (b) The inserts, having active electrodes as the sidewalls, are fabricated by electroplating the planar electrodes and sidewalls, patterning the channel, and drilling the in-/outlet.}
    \label{fig_sup:fab_process}
\end{figure}

The planar electrodes were fabricated with the same channel height, substrate, and cap layer as the insert with no electrodes. The fabrication process was modified in order to integrate the metallic structures into the chips depending on the required total thickness. The working electrodes were patterned out of \SI{2/1}{\nano \metre} and \SI{25/10}{\nano \metre} thick evaporated gold/chromium layers. Other structures such as the access tracks, pads, and counter electrodes were realised using a \SI{5}{\micro \metre} thick electroplated gold layer. In the first step, the substrate was covered with the Cr/Au seed layer. A mould for electroplating was then defined using SU-8 photoresist and UV-lithography followed by gold electroplating. In the next step, the mould was stripped and an SU-8 cover layer for the working electrodes was spincoated and patterned using UV-lithography. The unprotected seed layer was etched and the cover layer was stripped. The channel structures were further processed in an identical manner as for the electrode-free insert to form the channel and bond it to the top layer. The fabrication process is schematically illustrated in Figure~\ref{fig_sup:fab_process}a.
        
Sidewall-electrode inserts were fabricated in a similar way as the electrode-free insert, except for two intermediate metallisation steps to introduce the sidewalls, as well as the connecting tracks and the contact pads. For this purpose, the substrate was covered with a Cr/Au seed layer (\SI{20/60}{\nano \metre}). For the first mould, an SU-8 layer was patterned on the substrate followed by \SI{5}{\micro \metre} gold electroplating in order to define the tracks and the contact pads. In the next step, a second mould was formed on top of the first layer using SU-8 for the total height of \SI{90}{\micro \metre} in order to define the sidewall electrodes. The sidewall electrodes were electroplated for the desired height of \SI{80}{\micro \metre} (\SI{30}{\micro \metre} width). The mould structures were stripped followed by seed layer etching. A \SI{90}{\micro \metre} thick SU-8 was spincoated for both encapsulation of the electrodes and structuring the microfluidic channel. The rest of the process was the same as the electrode-free insert. The process is depicted in Figure~\ref{fig_sup:fab_process}b.

\subsection{\ce{B1} field distribution at the sample region}

\begin{figure}[htpb]
    \centering
    \includegraphics[width=0.65\linewidth]{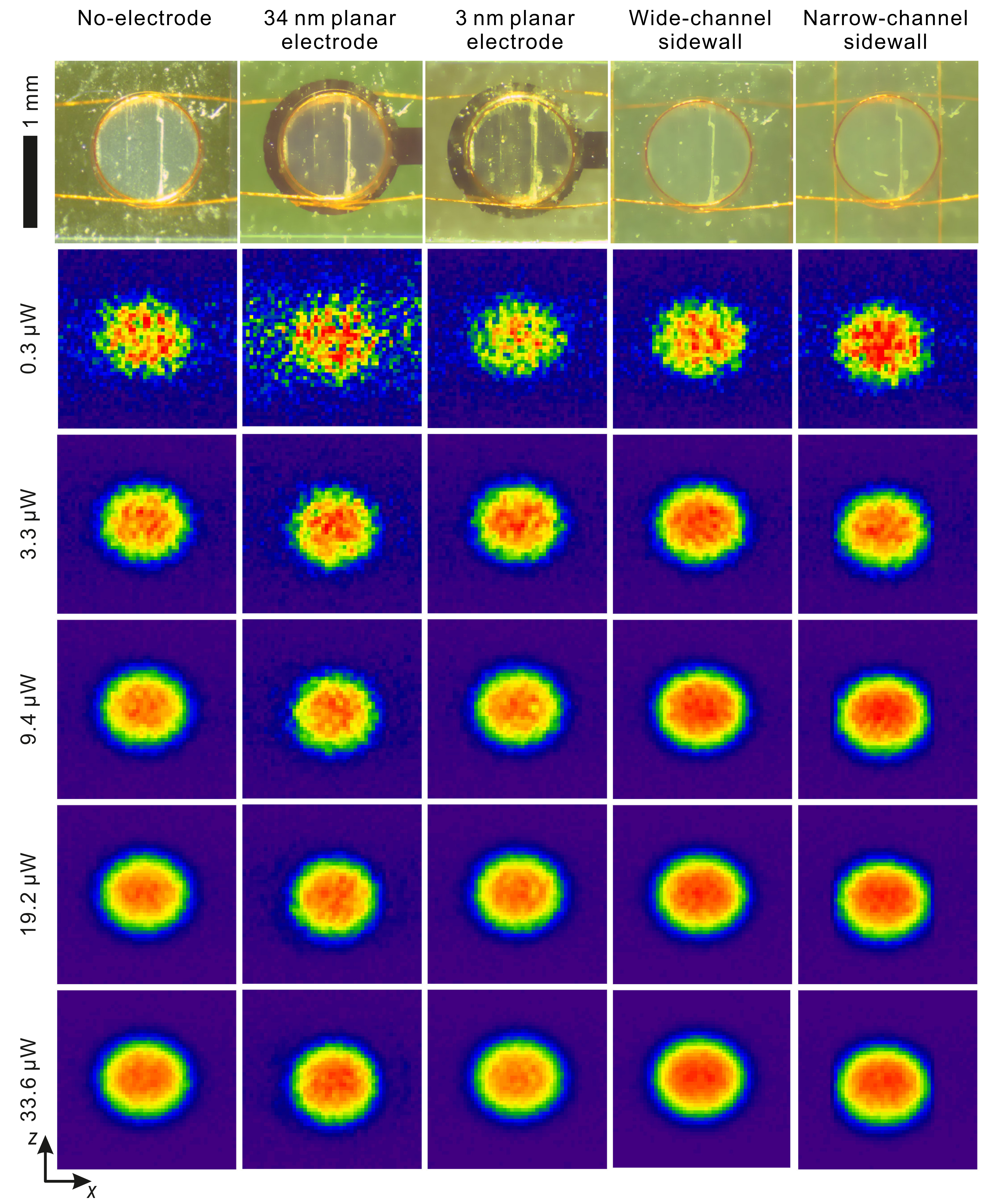}
    \caption{Distribution of the \ce{B1} field at the sample volume at different flip angles corresponding to a range of excitation powers. The asymmetric distribution in the $x$- and $z$-axis of the colour maps is due to the relative direction of the \ce{B0} and \ce{B1} fields: the \ce{B0} field is oriented along the $z$-axis, therefore the \ce{B1} components along this direction, i.e.,  parallel to \ce{B0}, do not produce any signal. }
    \label{fig_sup:B1map_MRI}
\end{figure}

%% file: main.bbl
\providecommand*{\mcitethebibliography}{\thebibliography}
\csname @ifundefined\endcsname{endmcitethebibliography}
{\let\endmcitethebibliography\endthebibliography}{}
\begin{mcitethebibliography}{40}
\providecommand*{\natexlab}[1]{#1}
\providecommand*{\mciteSetBstSublistMode}[1]{}
\providecommand*{\mciteSetBstMaxWidthForm}[2]{}
\providecommand*{\mciteBstWouldAddEndPuncttrue}
  {\def\EndOfBibitem{\unskip.}}
\providecommand*{\mciteBstWouldAddEndPunctfalse}
  {\let\EndOfBibitem\relax}
\providecommand*{\mciteSetBstMidEndSepPunct}[3]{}
\providecommand*{\mciteSetBstSublistLabelBeginEnd}[3]{}
\providecommand*{\EndOfBibitem}{}
\mciteSetBstSublistMode{f}
\mciteSetBstMaxWidthForm{subitem}
{(\emph{\alph{mcitesubitemcount}})}
\mciteSetBstSublistLabelBeginEnd{\mcitemaxwidthsubitemform\space}
{\relax}{\relax}

\bibitem[Badilita \emph{et~al.}(2012)Badilita, Meier, Spengler, Wallrabe, Utz,
  and Korvink]{Badilita2012}
V.~Badilita, R.~C. Meier, N.~Spengler, U.~Wallrabe, M.~Utz and J.~G. Korvink,
  \emph{Soft Matter}, 2012, \textbf{8}, 10583--10597\relax
\mciteBstWouldAddEndPuncttrue
\mciteSetBstMidEndSepPunct{\mcitedefaultmidpunct}
{\mcitedefaultendpunct}{\mcitedefaultseppunct}\relax
\EndOfBibitem
\bibitem[Fratila \emph{et~al.}(2014)Fratila, Gomez, S{\'{y}}kora, and
  Velders]{Fratila2014}
R.~M. Fratila, M.~V. Gomez, S.~S{\'{y}}kora and A.~H. Velders, \emph{Nature
  communications}, 2014, \textbf{5}, 3025\relax
\mciteBstWouldAddEndPuncttrue
\mciteSetBstMidEndSepPunct{\mcitedefaultmidpunct}
{\mcitedefaultendpunct}{\mcitedefaultseppunct}\relax
\EndOfBibitem
\bibitem[Montinaro \emph{et~al.}(2018)Montinaro, Grisi, Letizia, Pethö, Gijs,
  Guidetti, Michler, Brugger, and Boero]{Grisi2018}
E.~Montinaro, M.~Grisi, M.~C. Letizia, L.~Pethö, M.~A.~M. Gijs, R.~Guidetti,
  J.~Michler, J.~Brugger and G.~Boero, \emph{PLOS ONE}, 2018, \textbf{13},
  1--17\relax
\mciteBstWouldAddEndPuncttrue
\mciteSetBstMidEndSepPunct{\mcitedefaultmidpunct}
{\mcitedefaultendpunct}{\mcitedefaultseppunct}\relax
\EndOfBibitem
\bibitem[Gruschke \emph{et~al.}(2012)Gruschke, Baxan, Clad, Kratt, von
  Elverfeldt, Peter, Hennig, Badilita, Wallrabe, and Korvink]{Gruschke2012}
O.~G. Gruschke, N.~Baxan, L.~Clad, K.~Kratt, D.~von Elverfeldt, A.~Peter,
  J.~Hennig, V.~Badilita, U.~Wallrabe and J.~G. Korvink, \emph{Lab on a chip},
  2012, \textbf{12}, 495--502\relax
\mciteBstWouldAddEndPuncttrue
\mciteSetBstMidEndSepPunct{\mcitedefaultmidpunct}
{\mcitedefaultendpunct}{\mcitedefaultseppunct}\relax
\EndOfBibitem
\bibitem[van Meerten \emph{et~al.}(2018)van Meerten, Tijssen, van Bentum, and
  Kentgens]{VanMeerten2018}
S.~G. van Meerten, K.~C. Tijssen, P.~J. van Bentum and A.~P. Kentgens,
  \emph{Journal of Magnetic Resonance}, 2018, \textbf{286}, 60--67\relax
\mciteBstWouldAddEndPuncttrue
\mciteSetBstMidEndSepPunct{\mcitedefaultmidpunct}
{\mcitedefaultendpunct}{\mcitedefaultseppunct}\relax
\EndOfBibitem
\bibitem[{Oosthoek-de~Vries} \emph{et~al.}(2017){Oosthoek-de~Vries}, Bart,
  Tiggelaar, Janssen, van Bentum, Gardeniers, and
  Kentgens]{Oosthoek-deVries2017}
A.~J. {Oosthoek-de~Vries}, J.~Bart, R.~M. Tiggelaar, J.~W.~G. Janssen, P.~J.~M.
  van Bentum, H.~J. G.~E. Gardeniers and A.~P.~M. Kentgens, \emph{Analytical
  Chemistry}, 2017, \textbf{89}, 2296--2303\relax
\mciteBstWouldAddEndPuncttrue
\mciteSetBstMidEndSepPunct{\mcitedefaultmidpunct}
{\mcitedefaultendpunct}{\mcitedefaultseppunct}\relax
\EndOfBibitem
\bibitem[Hale \emph{et~al.}(2018)Hale, Rossetto, Greenhalgh, Finch, and
  Utz]{Hale2018}
W.~Hale, G.~Rossetto, R.~Greenhalgh, G.~Finch and M.~Utz, \emph{Lab Chip},
  2018, \textbf{18}, 3018--3024\relax
\mciteBstWouldAddEndPuncttrue
\mciteSetBstMidEndSepPunct{\mcitedefaultmidpunct}
{\mcitedefaultendpunct}{\mcitedefaultseppunct}\relax
\EndOfBibitem
\bibitem[Sharma and Utz(2019)]{Sharma2019}
M.~Sharma and M.~Utz, \emph{Journal of Magnetic Resonance}, 2019, \textbf{303},
  75 -- 81\relax
\mciteBstWouldAddEndPuncttrue
\mciteSetBstMidEndSepPunct{\mcitedefaultmidpunct}
{\mcitedefaultendpunct}{\mcitedefaultseppunct}\relax
\EndOfBibitem
\bibitem[Badilita \emph{et~al.}(2010)Badilita, Kratt, Baxan, Mohmmadzadeh,
  Burger, Weber, Elverfeldt, Hennig, Korvink, and Wallrabe]{Badilita2010}
V.~Badilita, K.~Kratt, N.~Baxan, M.~Mohmmadzadeh, T.~Burger, H.~Weber, D.~V.
  Elverfeldt, J.~Hennig, J.~G. Korvink and U.~Wallrabe, \emph{Lab on a chip},
  2010, \textbf{10}, 1387--1390\relax
\mciteBstWouldAddEndPuncttrue
\mciteSetBstMidEndSepPunct{\mcitedefaultmidpunct}
{\mcitedefaultendpunct}{\mcitedefaultseppunct}\relax
\EndOfBibitem
\bibitem[Meier \emph{et~al.}(2014)Meier, H{\"{o}}fflin, Badilita, Wallrabe, and
  Korvink]{Meier2014}
R.~C. Meier, J.~H{\"{o}}fflin, V.~Badilita, U.~Wallrabe and J.~G. Korvink,
  \emph{Journal of Micromechanics and Microengineering}, 2014, \textbf{24},
  045021\relax
\mciteBstWouldAddEndPuncttrue
\mciteSetBstMidEndSepPunct{\mcitedefaultmidpunct}
{\mcitedefaultendpunct}{\mcitedefaultseppunct}\relax
\EndOfBibitem
\bibitem[Spengler \emph{et~al.}(2014)Spengler, Moazenzadeh, Meier, Badilita,
  Korvink, and Wallrabe]{Spengler2014}
N.~Spengler, A.~Moazenzadeh, R.~Meier, V.~Badilita, J.~Korvink and U.~Wallrabe,
  \emph{Journal of Micromechanics and Microengineering}, 2014, \textbf{24},
  year\relax
\mciteBstWouldAddEndPuncttrue
\mciteSetBstMidEndSepPunct{\mcitedefaultmidpunct}
{\mcitedefaultendpunct}{\mcitedefaultseppunct}\relax
\EndOfBibitem
\bibitem[Spengler \emph{et~al.}(2016)Spengler, H{\"o}fflin, Moazenzadeh, Mager,
  MacKinnon, Badilita, Wallrabe, and Korvink]{Spengler2016}
N.~Spengler, J.~H{\"o}fflin, A.~Moazenzadeh, D.~Mager, N.~MacKinnon,
  V.~Badilita, U.~Wallrabe and J.~G. Korvink, \emph{PloS one}, 2016,
  \textbf{11}, e0146384\relax
\mciteBstWouldAddEndPuncttrue
\mciteSetBstMidEndSepPunct{\mcitedefaultmidpunct}
{\mcitedefaultendpunct}{\mcitedefaultseppunct}\relax
\EndOfBibitem
\bibitem[Bordonali \emph{et~al.}(2019)Bordonali, Nordin, Fuhrer, Mackinnon, and
  Korvink]{Bordonali2019}
L.~Bordonali, N.~Nordin, E.~Fuhrer, N.~Mackinnon and J.~G. Korvink, \emph{Lab
  on a Chip}, 2019, \textbf{19}, 503--512\relax
\mciteBstWouldAddEndPuncttrue
\mciteSetBstMidEndSepPunct{\mcitedefaultmidpunct}
{\mcitedefaultendpunct}{\mcitedefaultseppunct}\relax
\EndOfBibitem
\bibitem[Eills \emph{et~al.}(2019)Eills, Hale, Sharma, Rossetto, Levitt, and
  Utz]{Eills_High_2019}
J.~Eills, W.~Hale, M.~Sharma, M.~Rossetto, M.~H. Levitt and M.~Utz, \emph{J Am
  Chem Soc}, 2019, \textbf{141}, 9955--9963\relax
\mciteBstWouldAddEndPuncttrue
\mciteSetBstMidEndSepPunct{\mcitedefaultmidpunct}
{\mcitedefaultendpunct}{\mcitedefaultseppunct}\relax
\EndOfBibitem
\bibitem[Causier \emph{et~al.}(2015)Causier, Carret, Boutin, Berthelot, and
  Berthault]{Causier2015}
A.~Causier, G.~Carret, C.~Boutin, T.~Berthelot and P.~Berthault, \emph{Lab on a
  Chip}, 2015, \textbf{15}, 2049--2054\relax
\mciteBstWouldAddEndPuncttrue
\mciteSetBstMidEndSepPunct{\mcitedefaultmidpunct}
{\mcitedefaultendpunct}{\mcitedefaultseppunct}\relax
\EndOfBibitem
\bibitem[Nordin \emph{et~al.}(2019)Nordin, Bordonali, Badilita, and
  MacKinnon]{Nordin2019}
N.~Nordin, L.~Bordonali, V.~Badilita and N.~MacKinnon, \emph{Macromolecular
  bioscience}, 2019, \textbf{19}, 1800372\relax
\mciteBstWouldAddEndPuncttrue
\mciteSetBstMidEndSepPunct{\mcitedefaultmidpunct}
{\mcitedefaultendpunct}{\mcitedefaultseppunct}\relax
\EndOfBibitem
\bibitem[Richards and Evans(1975)]{Richards1975}
J.~A. Richards and D.~H. Evans, \emph{Analytical Chemistry}, 1975, \textbf{47},
  964--966\relax
\mciteBstWouldAddEndPuncttrue
\mciteSetBstMidEndSepPunct{\mcitedefaultmidpunct}
{\mcitedefaultendpunct}{\mcitedefaultseppunct}\relax
\EndOfBibitem
\bibitem[Mincey \emph{et~al.}(1990)Mincey, Popovich, Faustino, Hurst, and
  Caruso]{Mincey1990}
D.~W. Mincey, M.~J. Popovich, P.~J. Faustino, M.~M. Hurst and J.~A. Caruso,
  \emph{Analytical Chemistry}, 1990, \textbf{62}, 1197--1200\relax
\mciteBstWouldAddEndPuncttrue
\mciteSetBstMidEndSepPunct{\mcitedefaultmidpunct}
{\mcitedefaultendpunct}{\mcitedefaultseppunct}\relax
\EndOfBibitem
\bibitem[Prenzler \emph{et~al.}(2000)Prenzler, Bramley, Downing, and
  Heath]{Prenzler2000}
P.~D. Prenzler, R.~Bramley, S.~R. Downing and G.~A. Heath,
  \emph{Electrochemistry Communications}, 2000, \textbf{2}, 516--521\relax
\mciteBstWouldAddEndPuncttrue
\mciteSetBstMidEndSepPunct{\mcitedefaultmidpunct}
{\mcitedefaultendpunct}{\mcitedefaultseppunct}\relax
\EndOfBibitem
\bibitem[Albert \emph{et~al.}(1987)Albert, Dreher, Straub, and
  Rieker]{Albert1987}
K.~Albert, E.-L. Dreher, H.~Straub and A.~Rieker, \emph{Magnetic Resonance in
  Chemistry}, 1987, \textbf{25}, 919--922\relax
\mciteBstWouldAddEndPuncttrue
\mciteSetBstMidEndSepPunct{\mcitedefaultmidpunct}
{\mcitedefaultendpunct}{\mcitedefaultseppunct}\relax
\EndOfBibitem
\bibitem[Mairanovsky \emph{et~al.}(1983)Mairanovsky, Yusefovich, and
  Filippova]{Mairanovsky1983}
V.~G. Mairanovsky, L.~Y. Yusefovich and T.~M. Filippova, \emph{Journal of
  Magnetic Resonance (1969)}, 1983, \textbf{54}, 19--35\relax
\mciteBstWouldAddEndPuncttrue
\mciteSetBstMidEndSepPunct{\mcitedefaultmidpunct}
{\mcitedefaultendpunct}{\mcitedefaultseppunct}\relax
\EndOfBibitem
\bibitem[Webster(2004)]{Webster2004}
R.~D. Webster, \emph{Analytical Chemistry}, 2004, \textbf{76}, 1603--1610\relax
\mciteBstWouldAddEndPuncttrue
\mciteSetBstMidEndSepPunct{\mcitedefaultmidpunct}
{\mcitedefaultendpunct}{\mcitedefaultseppunct}\relax
\EndOfBibitem
\bibitem[Ni \emph{et~al.}(2017)Ni, Cui, Cao, and Chen]{Chen2017}
Z.-R. Ni, X.-H. Cui, S.-H. Cao and Z.~Chen, \emph{AIP Advances}, 2017,
  \textbf{7}, 085205\relax
\mciteBstWouldAddEndPuncttrue
\mciteSetBstMidEndSepPunct{\mcitedefaultmidpunct}
{\mcitedefaultendpunct}{\mcitedefaultseppunct}\relax
\EndOfBibitem
\bibitem[Ryan \emph{et~al.}(2014)Ryan, Smith, and Utz]{Ryan2014}
H.~Ryan, A.~Smith and M.~Utz, \emph{Lab Chip}, 2014, \textbf{14},
  1678--1685\relax
\mciteBstWouldAddEndPuncttrue
\mciteSetBstMidEndSepPunct{\mcitedefaultmidpunct}
{\mcitedefaultendpunct}{\mcitedefaultseppunct}\relax
\EndOfBibitem
\bibitem[Swyer \emph{et~al.}(2016)Swyer, Soong, Dryden, Fey, Maas, Simpson, and
  Wheeler]{Swyer_2016}
I.~Swyer, R.~Soong, M.~D. Dryden, M.~Fey, W.~E. Maas, A.~Simpson and A.~R.
  Wheeler, \emph{Lab on a Chip}, 2016, \textbf{16}, 4424--4435\relax
\mciteBstWouldAddEndPuncttrue
\mciteSetBstMidEndSepPunct{\mcitedefaultmidpunct}
{\mcitedefaultendpunct}{\mcitedefaultseppunct}\relax
\EndOfBibitem
\bibitem[Swyer \emph{et~al.}(2019)Swyer, von~der Ecken, Wu, Jenne, Soong,
  Vincent, Schmidig, Frei, Busse, Stronks, Simpson, and Wheeler]{Swyer_2019}
I.~Swyer, S.~von~der Ecken, B.~Wu, A.~Jenne, R.~Soong, F.~Vincent, D.~Schmidig,
  T.~Frei, F.~Busse, H.~J. Stronks, A.~J. Simpson and A.~R. Wheeler, \emph{Lab
  on a Chip}, 2019,  641--653\relax
\mciteBstWouldAddEndPuncttrue
\mciteSetBstMidEndSepPunct{\mcitedefaultmidpunct}
{\mcitedefaultendpunct}{\mcitedefaultseppunct}\relax
\EndOfBibitem
\bibitem[Wu \emph{et~al.}(2019)Wu, Ecken, Swyer, Li, Jenne, Vincent, Schmidig,
  Kuehn, Beck, Busse, Stronks, Soong, Wheeler, and Simpson]{Wu_2019}
B.~Wu, S.~Ecken, I.~Swyer, C.~Li, A.~Jenne, F.~Vincent, D.~Schmidig, T.~Kuehn,
  A.~Beck, F.~Busse, H.~Stronks, R.~Soong, A.~R. Wheeler and A.~Simpson,
  \emph{Angewandte Chemie Int Ed}, 2019, \textbf{58}, 15372--15376\relax
\mciteBstWouldAddEndPuncttrue
\mciteSetBstMidEndSepPunct{\mcitedefaultmidpunct}
{\mcitedefaultendpunct}{\mcitedefaultseppunct}\relax
\EndOfBibitem
\bibitem[Cheng \emph{et~al.}(2010)Cheng, Luo, Betz, Buckhout-White, Bekdash,
  Payne, Bentley, and Rubloff]{cheng2010}
Y.~Cheng, X.~Luo, J.~Betz, S.~Buckhout-White, O.~Bekdash, G.~F. Payne, W.~E.
  Bentley and G.~W. Rubloff, \emph{Soft Matter}, 2010, \textbf{6},
  3177--3183\relax
\mciteBstWouldAddEndPuncttrue
\mciteSetBstMidEndSepPunct{\mcitedefaultmidpunct}
{\mcitedefaultendpunct}{\mcitedefaultseppunct}\relax
\EndOfBibitem
\bibitem[Li \emph{et~al.}(2018)Li, Liu, Kim, Song, Tsao, Teng, Gao, Mei,
  Bentley, Payne,\emph{et~al.}]{li2018}
Y.~Li, Y.~Liu, E.~Kim, Y.~Song, C.-Y. Tsao, Z.~Teng, T.~Gao, L.~Mei, W.~E.
  Bentley, G.~F. Payne \emph{et~al.}, \emph{Carbohydrate polymers}, 2018,
  \textbf{195}, 505--514\relax
\mciteBstWouldAddEndPuncttrue
\mciteSetBstMidEndSepPunct{\mcitedefaultmidpunct}
{\mcitedefaultendpunct}{\mcitedefaultseppunct}\relax
\EndOfBibitem
\bibitem[Zhuang \emph{et~al.}(2018)Zhuang, Lin, Dong, Cheng, and
  Weng]{zhuang2018}
J.~Zhuang, S.~Lin, L.~Dong, K.~Cheng and W.~Weng, \emph{ACS Biomaterials
  Science \& Engineering}, 2018, \textbf{4}, 1528--1535\relax
\mciteBstWouldAddEndPuncttrue
\mciteSetBstMidEndSepPunct{\mcitedefaultmidpunct}
{\mcitedefaultendpunct}{\mcitedefaultseppunct}\relax
\EndOfBibitem
\bibitem[Chung \emph{et~al.}(2018)Chung, Chandra, Koo, and Shim]{chung2018}
S.~Chung, P.~Chandra, J.~P. Koo and Y.-B. Shim, \emph{Biosensors and
  Bioelectronics}, 2018, \textbf{100}, 396--403\relax
\mciteBstWouldAddEndPuncttrue
\mciteSetBstMidEndSepPunct{\mcitedefaultmidpunct}
{\mcitedefaultendpunct}{\mcitedefaultseppunct}\relax
\EndOfBibitem
\bibitem[Schenck(1996)]{Schenck1996}
J.~F. Schenck, \emph{Medical Physics}, 1996, \textbf{23}, 815--850\relax
\mciteBstWouldAddEndPuncttrue
\mciteSetBstMidEndSepPunct{\mcitedefaultmidpunct}
{\mcitedefaultendpunct}{\mcitedefaultseppunct}\relax
\EndOfBibitem
\bibitem[Wapler \emph{et~al.}(2014)Wapler, Leupold, Dragonu, von Elverfeld,
  Zaitsev, and Wallrabe]{Wapler2014}
M.~C. Wapler, J.~Leupold, I.~Dragonu, D.~von Elverfeld, M.~Zaitsev and
  U.~Wallrabe, \emph{Journal of Magnetic Resonance}, 2014, \textbf{242}, 233 --
  242\relax
\mciteBstWouldAddEndPuncttrue
\mciteSetBstMidEndSepPunct{\mcitedefaultmidpunct}
{\mcitedefaultendpunct}{\mcitedefaultseppunct}\relax
\EndOfBibitem
\bibitem[Milazzo \emph{et~al.}(1978)Milazzo, Caroli, and Braun]{Milazzo1978}
G.~Milazzo, S.~Caroli and R.~D. Braun, \emph{Journal of The Electrochemical
  Society}, 1978, \textbf{125}, 261C\relax
\mciteBstWouldAddEndPuncttrue
\mciteSetBstMidEndSepPunct{\mcitedefaultmidpunct}
{\mcitedefaultendpunct}{\mcitedefaultseppunct}\relax
\EndOfBibitem
\bibitem[Bard(2017)]{Bard2017}
A.~Bard, \emph{{Standard potentials in aqueous solution}}, Routledge,
  2017\relax
\mciteBstWouldAddEndPuncttrue
\mciteSetBstMidEndSepPunct{\mcitedefaultmidpunct}
{\mcitedefaultendpunct}{\mcitedefaultseppunct}\relax
\EndOfBibitem
\bibitem[Geng \emph{et~al.}(2016)Geng, Wang, Guo, Zhang, Chen, and
  Wang]{Geng2016}
Z.~Geng, X.~Wang, X.~Guo, Z.~Zhang, Y.~Chen and Y.~Wang, \emph{Journal of
  Materials Chemistry B}, 2016, \textbf{4}, 3331--3338\relax
\mciteBstWouldAddEndPuncttrue
\mciteSetBstMidEndSepPunct{\mcitedefaultmidpunct}
{\mcitedefaultendpunct}{\mcitedefaultseppunct}\relax
\EndOfBibitem
\bibitem[Webb(2013)]{Webb2013}
A.~Webb, \emph{Journal of Magnetic Resonance}, 2013, \textbf{229}, 55--66\relax
\mciteBstWouldAddEndPuncttrue
\mciteSetBstMidEndSepPunct{\mcitedefaultmidpunct}
{\mcitedefaultendpunct}{\mcitedefaultseppunct}\relax
\EndOfBibitem
\bibitem[da~Silva \emph{et~al.}(2019)da~Silva, Gomes, Lobo, J{\'u}nior,
  Danieli, Carmo, Bl{\"u}mich, and Colnago]{da_Silva_Electrochemical_2019}
P.~da~Silva, B.~Gomes, C.~Lobo, L.~J{\'u}nior, E.~Danieli, M.~Carmo,
  B.~Bl{\"u}mich and L.~Colnago, \emph{Microchem J}, 2019, \textbf{146},
  658--663\relax
\mciteBstWouldAddEndPuncttrue
\mciteSetBstMidEndSepPunct{\mcitedefaultmidpunct}
{\mcitedefaultendpunct}{\mcitedefaultseppunct}\relax
\EndOfBibitem
\bibitem[Gomes \emph{et~al.}(2017)Gomes, da~Silva, Lobo, da~Santos, and
  Colnago]{Gomes_Strong_2017}
B.~Gomes, P.~da~Silva, C.~Lobo, M.~da~Santos and L.~Colnago, \emph{Anal Chim
  Acta}, 2017, \textbf{983}, 91--95\relax
\mciteBstWouldAddEndPuncttrue
\mciteSetBstMidEndSepPunct{\mcitedefaultmidpunct}
{\mcitedefaultendpunct}{\mcitedefaultseppunct}\relax
\EndOfBibitem
\bibitem[Ni \emph{et~al.}(2017)Ni, Cui, Cao, and Chen]{Ni_A_2017}
Z.~Ni, X.~Cui, S.~Cao and Z.~Chen, \emph{Aip Adv}, 2017, \textbf{7},
  085205\relax
\mciteBstWouldAddEndPuncttrue
\mciteSetBstMidEndSepPunct{\mcitedefaultmidpunct}
{\mcitedefaultendpunct}{\mcitedefaultseppunct}\relax
\EndOfBibitem
\end{mcitethebibliography}
